\documentclass[11pt,one column]{IEEEtran} 

\usepackage{graphics}
\usepackage{subfig}
\usepackage{color}
\usepackage{epsfig}
\usepackage{amssymb,amsmath,amsthm}
\usepackage{latexsym}
\usepackage{mathrsfs}
\usepackage{bbm, ulem}
 \usepackage{setspace, hyperref}
\usepackage{epstopdf, titling}
\usepackage{mathtools}
\usepackage[hang,flushmargin]{footmisc}
 \usepackage{flushend}
 \usepackage{fancybox}
\usepackage{relsize}

\mathtoolsset{showonlyrefs}

\newcommand{\cX}{{\mathcal X}}
\newcommand{\cY}{{\mathcal Y}}

\newcommand{\cB}{{\mathcal B}}

\newcommand{\cM}{{\mathcal M}}
\newcommand{\cU}{{\mathcal U}}
\newcommand{\cJ}{{\mathcal J}}
\newcommand{\cP}{{\mathcal P}}
\newcommand{\cAk}{{\mathcal A}_{k}}

\newcommand{\cXm}{{\mathcal X}_{\mathcal M}}
\newcommand{\cYm}{{\mathcal Y}_{\mathcal M}}

\newcommand{\cYB}{{\mathcal Y}_{B}}

\newcommand{\ep}{\epsilon}

\newcommand\MC{{ \ - \!\!\circ\!\! - \ }}

\newtheorem{theorem}{Theorem}
\newtheorem{lemma}[theorem]{Lemma}
\newtheorem{proposition}[theorem]{Proposition}
\newtheorem*{corollary}{Corollary}

\renewcommand*{\thefootnote}{\fnsymbol{footnote}}
\theoremstyle{remark}
\newtheorem*{remark*}{Remark}
\newtheorem*{remarks*}{Remarks}

\theoremstyle{definition}
\newtheorem{definition}{Definition}

\newtheorem{example}{Example}

\allowdisplaybreaks

\begin{document}

\date{}
\title{ Universal  Sampling Rate Distortion}
\author{Vinay Praneeth Boda and Prakash Narayan$^\dag$ }
\maketitle 

\renewcommand{\figurename}{Figure}

{\renewcommand{\thefootnote}{} \footnotetext{

\noindent 
$^\dag$V.~P. Boda and P. Narayan are with the Department of
Electrical and Computer Engineering and the Institute for Systems
Research, University of Maryland, College Park, MD 20742, USA.
E-mail: \{praneeth, prakash\}@umd.edu.

Part of this work was presented at the International Symposium
on Information Theory, Aachen, Germany,  2017.

\noindent This work was supported by the U.S.
National Science Foundation under Grants CCF-0917057 and CCF-1319799.}}

\thispagestyle{plain}
\pagestyle{plain}

\vspace*{0.1cm}

\begin{abstract}

We examine the coordinated and universal rate-efficient sampling of a 
subset of correlated discrete memoryless sources followed by lossy compression
of the sampled sources. The goal is to reconstruct a predesignated subset of sources 
within a specified level of distortion. 
The combined sampling mechanism and rate distortion code are universal in that 
they are devised to perform robustly without exact knowledge of 
the underlying joint probability distribution of the sources. In Bayesian as well as
nonBayesian settings, single-letter characterizations are provided for the universal sampling 
rate distortion function for fixed-set sampling, independent random sampling 
and memoryless random sampling. It is illustrated how these sampling mechanisms are successively
better. Our achievability proofs bring forth new schemes for joint source 
distribution-learning and lossy compression.
\end{abstract}

\vspace*{0.05cm}




\section{Introduction}

Consider a set $\cM$ of $m$ discrete memoryless sources with joint probability mass 
function (pmf)  known only to belong to a given family of pmfs.
At time instants $t = 1, \ldots,n,$ 
possibly different subsets $  A_t$ of size $  k \leq m$ are sampled ``spatially'' 
and compressed jointly by a (block) source code, with the objective of reconstructing a predesignated subset
$B \subseteq \cM$ of sources from the compressed representations  within a specified level of distortion.
In forming  an efficient rate distortion code that
yields the best compression rate for a given distortion level,
what are the tradeoffs -- under optimal 
processing -- among {\it causal} sampling procedure, inferential methods for
approximating the underlying joint pmf of the memoryless sources, compression rate and distortion level?
 ``Universality''  requires
that the combined sampling mechanism
and lossy compression code  be fashioned in the face of imprecise
knowledge of the underlying pmf. 
This paper is a progression of our work in \cite{BodNar17}
on sampling rate distortion for multiple sources with {\it known} joint pmf.
Motivating applications include in-network 
computation \cite{GiriKum05}, dynamic thermal management in multicore
processor chips \cite{ZhaSri10}, etc.

\vspace*{0.1cm}

The study of problems of combined sampling and compression has a 
classical and distinguished history.
Recent relevant works include the lossless compression
of analog sources in an information theoretic setting \cite{WuVer10};
compressed sensing with an allowed detection error rate or quantization
distortion \cite{ReeGas12}; sub-Nyquist temporal sampling followed
by lossy reconstruction \cite{KipnisGold16}; and rate distortion function for 
multiple sources with time-shared sampling \cite{LiuSimErk12}.
See also \cite{IshKunRam03, WeiVet12}.
Closer to our approach that entails {\it spatial} sampling, 
the rate distortion function has 
been characterized when multiple Gaussian signals from a random 
field are sampled and quantized (centralized or distributed) in 
\cite{NeuPra11}. 
In a setting of distributed acoustic 
sensing and reconstruction, centralized as well as distributed 
coding schemes and sampling lattices are studied  in 
\cite{KonTelVet12}.
In \cite{KashLasXiaLiu05}, a 
Gaussian random field on the interval $[0,1]$ and i.i.d. 
in time, is reconstructed  from compressed 
versions of $k$ sampled sequences under a mean-squared error 
distortion criterion. 
All the sampling problems  above assume a knowledge of
the underlying pmf. 
 
\vspace*{0.1cm}

In the realm of rate distortion theory where a complete knowledge of
the signal statistics is unknown, there is a rich literature that considers
various formulations of universal coding; only a sampling is listed here. 
Directions include 
classical Bayesian and nonBayesian methods \cite{Ziv72, NeuGrDav75, NeuShi78, Rissanen84};
``individual sequences'' studies \cite{Ziv80, WeissMer01, WeissMer02a};
redundancy in quantization rate or distortion \cite{LinLugZeg95, Linder00, Linder02};
and lossy compression of noisy or remote signals \cite{LinLugZeg97, Weiss01, DemboWeiss03}. 
These works propose a variety of distortion measures to investigate
universal reconstruction performance. 

\vspace*{0.1cm}

Our work differs materially from the approaches above. Sampling is
spatial rather than temporal, unlike in most of the settings above.
Furthermore, we introduce new forms of 
randomized sampling that can depend on the observed source realizations,
and which yield a clear gain in performance.
We restrict ourselves to universality that involves a lack
of specific knowledge of source pmf within a {\it finite}
family of pmfs. Accordingly, in Bayesian and nonBayesian settings,
we consider average and peak distortion criteria, respectively,
with an emphasis on the former. Extensions to
an infinite family of pmfs are currently under study.

\vspace*{0.1cm}

Our technical contributions are as follows. In Bayesian and nonBayesian settings,
we consider a new formulation involving
an {\it universal sampling rate distortion function} (USRDf), with the objective
of capturing the interplay and characterizing inherent tradeoffs among 
sampling mechanism, approximation of underlying (unknown) pmf, 
lossy compression rate and distortion level. Our results build on
the concept of sampling rate distortion function \cite{BodNar17}, which in turn 
uses as an ingredient the rate distortion function for a ``remote'' source-receiver 
model with known pmf \cite{DobTsy62, Ber71, Ber78, YamIto80}. 
We begin with fixed-set sampling where  
the encoder observes the same set of $k$ sampled sources at every time instant. 
Recognizing that only the $k$-marginal
pmf of the sources -- pertaining to the sampling set -- can be learned by the 
encoder, the corresponding USRDf is characterized. 
In general, allowing randomization in sampling affords two distinct 
advantages over fixed-set sampling: 
better approximation of the underlying joint pmf and improved compression
performance enabled by sampling different subsets of sources 
in apposite proportions. An {\it independent} random sampler chooses 
different $k$-subsets of the sources independently of source realizations and
independently in time, and can learn all $k$-marginals of the joint pmf.
This reduction in pmf uncertainty (vis-{\`a}-vis fixed-set sampling)
aids in improving USRDf. Interestingly, our achievability proof
shows how this USRDf can be attained without informing the decoder 
{\it explicitly} of the sampling sequence. Lastly, we consider
a more powerful sampler, namely the memoryless random sampler, whose
choice of sampling sets can depend on instantaneous source realizations. 
Surprisingly, this latitude allows the encoder to learn the entire 
joint pmf, and that, too, only from the sampling sequence without recourse to the sampled
source realizations. 
Furthermore, we show how USRDf can be attained by means of a sampling sequence
that depends {\it deterministically} on source realizations, thereby reducing
code complexity. Thus, all our achievability proofs bring out new ideas for joint
source pmf-learning and lossy compression.

\vspace*{0.1cm}

Our model is described in Section \ref{s:Preliminaries}. The main results,
illustrated by examples, are stated in Section \ref{s:Results}. In 
Section \ref{s:Proofs}, we present the achievability proofs in the increasing
order of sampler complexity, with an emphasis on the Bayesian setting; a unified
converse proof is presented thereafter.



\vspace*{-0.1cm}

\section{Preliminaries} \label{s:Preliminaries}

Denote $\cM = \{ 1, \ldots, m \}$, and let $X_{\cM} = ( X_{1}, \ldots,
X_{m} )$ be a $\cX_{\cM} = \mathop{\mbox{\large $\times$}} 
\limits_{i=1}^m \cX_{i}$-valued rv where each $\cX_{i}$ is a 
finite alphabet. For a (nonempty) set $A \subseteq \cM $, we denote by 
$X_{A}$ the rv $ ( X_{i}, i \in A ) $ with values in 
$\mathop{\mbox{\large $\times$}} \limits_{i \in A} \cX_{i}$, 
and denote $n$  repetitions of $X_{A}$ by $X_{A}^{n} 
= ( X_{i}^{n}, i \in A)$ with values in $\cX_{A}^{n} 
= \mathop{\mbox{\large $\times$}} \limits_{i \in A} \cX_{i}^{n}$, 
where $X_{i}^{n} = ( X_{i1}, \ldots, X_{in})$ takes values in 
the $n$-fold product space ${\cal X}_{i}^{n} = \cX_{i} 
\times \cdots \times \cX_{i}$. For $1 \leq k \leq m$, let 
${\cal A}_{k} = \{ A:  A \subseteq \cM, \ |A| = k \}$ be the set 
of all  $k$-sized subsets of $\cM$ and let 
$A^{c}  = \cM \! \setminus \! A$. Let $\cY_{\cM} = \mathop{\mbox{\large $\times$}} \limits_{i=1}^m \cY_{i},$ 
where $\cY_{i}$ is a finite reproduction alphabet for $X_{i}$.
All logarithms and exponentiations 
are with respect to the base 2.

\vspace*{0.1cm}

Let $\Theta$ be a finite set (of parameters) and $\theta$ a $\Theta$-valued rv with pmf
$\mu_{\theta}$ of assumed full support.
We consider a discrete memoryless multiple source (DMMS) $\{ X_{\cM t} 
\}_{t=1}^{\infty}$  consisting of i.i.d. repetitions of the rv 
$X_{\cM}$ with pmf known only to the extent of 
belonging to a finite family of pmfs $\cP = \{ P_{X_{\cM}|\theta = \tau}, \ \tau \in \Theta \}$
of assumed full support.
Two settings are studied: in a Bayesian formulation, the pmf $\mu_{\theta}$
is taken to be {\it known} while in a nonBayesian formulation $\theta$ 
is an {\it unknown constant} in $\Theta$.

\begin{definition}\label{d: RS}
In the Bayesian setting, a $k$-{\it random sampler} ($k$-RS), $1 \leq k \leq m$, collects 
 causally  at each $t=1, \ldots, n$, random samples
 $X_{S_{t}} \triangleq X_{S_{t} t}$ from $X_{\cM t}$, where $S_{t}$ is a rv with values in 
 $\cAk$ with (conditional) pmf $P_{S_{t} | X_{\cM}^{t} S^{t-1}} $, 
 with $X_{\cM}^{t} = (X_{\cM 1}, \ldots, X_{\cM t} )$ and $S^{t-1}
 =  ( S_{1}, \ldots , S_{t-1} ) $. Such a $k$-RS is specified by a 
 (conditional) pmf $P_{S^{n}|X_{\cM}^{n} \theta }$ with the requirement  
 \begin{equation}
  \label{eq:k-RS-distribution}
 P_{S^{n} | X_{\cM}^{n} \theta } = P_{S^{n} | X_{\cM}^{n}}
 = \prod_{t=1}^{n} P_{S_{t} | 
 X_{\cM}^{t} S^{t-1} }.      
 \vspace*{-0.1cm}
 \end{equation}
 In the nonBayesian setting, the first equality above is redundant.
 In both settings, a $k$-RS is unaware of the underlying pmf of the DMMS. 
 
\vspace*{0.2cm} 
 
The output of a $k$-RS is $(S^{n},X_{S }^{n})$ where $X_{S }^{n} 
= (X_{S_{1} },\ldots,X_{S_{n} })$. Successively restrictive choices 
of a $k$-RS in \eqref{eq:k-RS-distribution} corresponding to \vspace*{-0.1cm}
  \begin{equation}
 \label{eq:k_MRS_def}
  P_{S_{t}|X_{\cM}^{t} S^{t-1}} = P_{S_{t}|X_{\cM t}}, \ \ t=1,
  \ldots,n,
 \end{equation}
\begin{equation}
\label{eq:k_IRS_def}
 P_{S_{t}|X_{\cM}^{t} S^{t-1}} = P_{S_{t}}, \ \ \ \ \ \ \  t=1, 
 \ldots,n,
\end{equation}
and, for a given $A \subseteq \cM,$
\begin{equation}
\label{eq:k_FS}
 P_{S_{t}|X_{\cM}^{t} S^{t-1}} = \mathbbm{1}(S_{t} = A), \ \ \ \ \ \  t=1, 
 \ldots,n
\end{equation}
will be termed the $k$-{\it memoryless 
random sampler}, $k$-{\it independent 
random sampler} and the $k$-{\it fixed-set sampler} abbreviated as $k$-MRS, 
$k$-IRS and $k$-FS, respectively.
\end{definition}


\vspace*{-0.1cm}

Our objective is to reconstruct a subset 
of DMMS components with indices in an arbitrary but fixed
 {\it recovery set} $B \subseteq \cM$,
namely $X_{B}^{n},$ from a compressed representation of the $k$-RS output
$(S^{n}, X_{S}^{n}),$ under a suitable distortion criterion.

\begin{definition} \label{d:encoder}
An $n$-length block code with $k$-RS for a DMMS $\{ X_{\cM t}
 \}_{t=1}^{ \infty} $ with alphabet $\cX_{\cM}$ and reproduction 
 alphabet $\cY_{B}$ is a triple $(P_{S^{n}|X_{\cM}^{n}}, f_{n}, 
 \varphi_{n} )$ where $P_{S^{n} | X_{\cM}^{n} }$ is a $k$-RS as in 
 \eqref{eq:k-RS-distribution}, and  $(f_{n}, \varphi_{n} )$ are a 
 pair of mappings where the encoder $f_{n}$ maps the $k$-RS output 
 $(S^{n},X_{S}^{n})$ into some finite set $\cJ = \{1, \ldots , J \}$ 
 and the decoder $\varphi_{n}$, with access to $S^{n}$ and
 the encoder output, maps 
 $\cAk^{n} \times \cJ$ into $\cY_{B}^{n}$. We 
 shall use the compact notation $(P_{S|X_{\cM}}, f, \varphi),$ 
 suppressing  $n$. The rate of the code with $k$-RS 
 $(P_{S|X_{\cM}}, f, \varphi)$  is $\tfrac{1}{n} \log ||f|| =  \tfrac{1}{n}\log J$.
(An encoder that operates by forming first an explicit
estimate  of $\theta$ from $(S^{n},X_{S}^{n})$ is 
subsumed by this definition.)
 \end{definition}

\noindent {\it Remark}: We note that the decoder $\varphi$ above is 
taken to be informed of the sequence of sampling sets $S^{n}$. This assumption is meaningful
for a $k$-IRS and $k$-MRS. For a $k$-IRS, it will be shown to be not needed.

\vspace*{0.1cm}
For a given (single-letter) finite-valued distortion measure 
$d: \cX_{B} \times \cY_{B} \rightarrow \mathbb{R}^{+} \cup 
\{0\} $, an $n$-length block code with $k$-RS 
$(P_{S|X_{\cM}}, f, \varphi)$ will be required to satisfy 
one of the following distortion criteria $(d,\Delta)$ depending on the 
setting.

\noindent (i) Bayesian: The {\it expected} distortion criterion is
\begin{align}
\begin{split}
 \mathbb{E} \Big [   d \Big (  X_{B}^{n},  \varphi 
  \big (S^{n},  f (S^{n},  X_{S}^{n} ) \big ) \Big ) \Big ]  
 & \triangleq  \mathbb{E} \Big[ \dfrac{1}{n} \sum_{t=1}^{n} 
 d \Big ( X_{B t} , \Big ( \varphi \big ( S^{n}, 
 f ( S^{n}, X_{S}^{n}) \big ) \Big )_{t}  \Big ) \Big ]
   \\ 
 & = \sum \limits_{\tau \in \Theta} \mu_{\theta}(\tau) 
   \mathbb{E} \Big [ \dfrac{1}{n} \sum_{t=1}^{n} 
 d \Big ( X_{B t} , \Big ( \varphi \big (S^{n},  
 f ( S^{n}, X_{S}^{n}) \big ) \Big )_{t}  \Big ) \Big | \theta = \tau \Big ] \vspace*{-0.4cm} \\ 
 \label{eq:expected-distortion} 
 & \leq  \Delta.  
 \vspace*{-0.2cm}
 \end{split}
 \vspace*{-0.2cm}
\end{align}
(ii) NonBayesian: The {\it peak} distortion criterion is
\begin{align}
\begin{split}
 \label{eq:peak-distortion}
 \underset{\tau \in \Theta} \max \ \mathbbm{E} \Big [  d \Big (  X_{B}^{n},  \varphi 
  \big (S^{n},  f (S^{n},  X_{S}^{n} ) \big ) \Big ) \big | \theta = \tau \Big ] 
 \leq \Delta,
 \end{split}
\end{align}
where the ``conditional'' expectation denotes, in fact, $\mathbbm{E}_{P_{X_{\cM}^{n} S^{n}| \theta = \tau} } = 
\mathbbm{E}_{P_{X_{\cM}^{n} | \theta = \tau}  P_{S^{n}|X_{\cM}^n} }.$

\begin{definition} \label{d:RDF}
A number $ R \geq 0$ is an achievable universal $k$-RS coding rate at 
 distortion level $\Delta$ if for every $\epsilon > 0$ 
and sufficiently large $n$, there exist $n$-length block codes 
with $k$-RS of rate less than $R + \epsilon$ and satisfying the 
distortion criterion $(d, \Delta + \ep)$ in \eqref{eq:expected-distortion} or 
\eqref{eq:peak-distortion} above; and 
$(R, \Delta)$ will be termed an achievable universal $k$-RS 
rate distortion pair under the expected or peak distortion criterion.
The infimum of such achievable rates is 
denoted by $R_{A}( \Delta)$, $R_{\imath}( \Delta)$ and $R_{m}(\Delta)$ 
for a $k$-FS, $k$-IRS and  $k$-MRS, respectively. We shall refer to 
$R_{A}(\Delta), \ R_{\imath}(\Delta)$  as well as $R_{m}(\Delta)$ 
as the {\it universal sampling rate distortion function} (USRDf), 
suppressing the dependence on $k$.
\end{definition}

\noindent {\it Remark}: Clearly, the USRDf under \eqref{eq:expected-distortion}  will
be no larger than that under \eqref{eq:peak-distortion}. 
\ifx
(i) In the setting of an informed decoder,
the sampling mechanism has two means of conveying information
regarding $X_{\cM}^{n}$ to the decoder: via the encoder output as well
as by embedding it implicitly in $S^{n}$.

\vspace*{0.1cm}
\noindent (ii) Clearly, $R^{I}(\Delta) \leq 
R^{U}(\Delta),$ and both are nonincreasing in $k$.

\vspace*{0.1cm}

\noindent (iii) For a DMMS $\{ X_{\cM t} \}_{t=1}^{\infty}$, the requirement 
\eqref{eq:k_MRS_def} on the sampler renders $\left \{ (X_{\cM t}, 
S_{t}) \right \}_{t=1}^{\infty}$ and thereby also 
$\left \{ (X_{S_{t}}, S_{t}) \right \}_{t=1}^{\infty}$ to be 
memoryless sequences. 
\fi



\section{Main Results}\label{s:Results}

We make the following main contributions. First, a (single-letter) characterization is provided
of the USRDf  for fixed-set sampling, i.e., $k$-FS, in the Bayesian and nonBayesian settings.
Second, building on this, a characterization of the USRDf is obtained for
a $k$-IRS in these settings, and it is shown that randomized sampling can outperform
strictly the ``best'' fixed-set sampler. Indeed, this USRDf can be attained even upon
dispensing with the a priori assumption that the decoder is informed of the sequence of sampling
sets. Finally, the USRDf for a $k$-MRS is characterized and shown to be achievable by a sampler
that is determined by the instantaneous realizations of the DMMS at each time instant.
We note that the USRDfs for a $k$-FS and $k$-IRS can be 
deduced from that of a $k$-MRS. Nevertheless, for the sake of expository convenience, 
we develop the three sampling models in succession; this will also facilitate the presentation of the 
achievability proofs.
 
\vspace*{0.1cm}

Throughout this paper, a salient theme that recurs is this: An encoder without
prior knowledge of $\theta$ and with access to only $k$ instantaneously sampled components of the DMMS
$\{ X_{\cM t}\}_{t=1}^{\infty}$ can form only a limited estimate of $\theta.$ 
The quality of said estimate improves steadily from $k$-FS to $k$-IRS to $k$-MRS.

\vspace*{0.1cm}

Consider first fixed-set sampling with $A \subseteq \cM$  in \eqref{eq:k_FS}.
An encoder $f$ with access to $X_{A}^{n}$ cannot distinguish among 
pmfs in $\cP$ (indexed by $\tau$) that have the same $P_{X_{A}|\theta = \tau}.$
Accordingly, let  $\Theta_1$ be a partition of $\Theta$ comprising ``ambiguity''
atoms, with each such atom consisting of $\tau$s with {\it identical marginal pmfs} $P_{X_{A}|\theta = \tau}$.
Indexing the elements of $\Theta_1$ by $\tau_1$, let $\theta_1$ be a $\Theta_1$-valued
rv with pmf $\mu_{\theta_1}$ induced by $\mu_{\theta}.$ For each $\tau_1 \in \Theta_1,$
let $\Lambda(\tau_1)$ be the collection of $\tau$s in the atom of $\Theta_1$
indexed by $\tau_1$. In the Bayesian setting, clearly
\begin{align}
\label{eq:Lambda_def}
  P_{X_{A}|\theta_1 = \tau_1} = P_{X_{A}|\theta = \tau}, \ \ \ \tau \in \Lambda(\tau_1).
\end{align}
In the nonBayesian setting, in order to retain the same notation, we choose 
$P_{X_{A}|\theta_1 = \tau_1}$ to be the right-side above.

\begin{figure}[h]
\begin{center}
    \includegraphics[width=7cm,height=5cm]{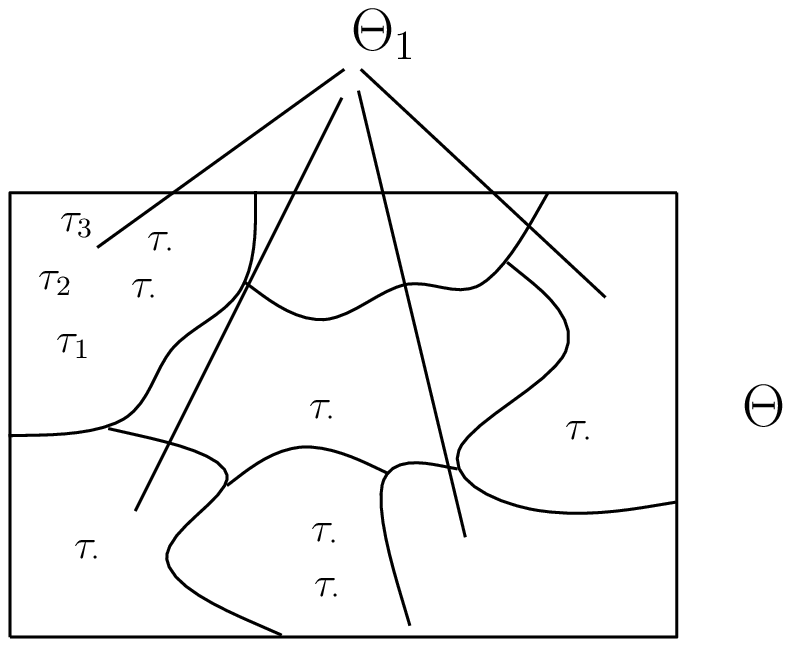}  
\end{center}  
     \caption{Ambiguity atoms}
  \label{fig:IRS_Fixed_set}
\end{figure}

When the pmf  of the DMMS $\{X_{\cM t} \}_{t=1}^{\infty}$ is {\it known}, say 
$P_{X_{\cM}}$  -- corresponding to $|\Theta| = 1$ -- we recall from \cite{BodNar17}  
that the (U)SRDf for fixed $A \subseteq \cM $ is
\begin{align} \label{eq:FS_SRDf}
 R_{A}(\Delta) = \underset{ X_{\cM} \MC X_{A} \MC Y_{B} \atop \mathbbm{E}[d(X_{B}, Y_{B})] 
 \leq \Delta} \min I(X_{A} \wedge Y_{B}), \ \ \ \ \Delta_{\min} \leq \Delta \leq \Delta_{\max},
\end{align}
with 
\begin{align}
 \Delta_{\min} = \mathbbm{E}\big [ \underset{ y_{B} \in \cY_{B} } \min 
 \mathbbm{E}[d(X_{B}, y_{B})|X_{A}]  \big ],
 \quad 
 \Delta_{\max} =   \underset{ y_{B} \in \cY_{B} } \min \big 
 [ \mathbbm{E}[d(X_{B}, y_{B})|X_{A}]  \big ],
\end{align}
which can be interpreted as the (standard) rate distortion function 
for the DMMS $\{ X_{A t} \}_{t=1}^{\infty}$ using a modified distortion measure ${\tilde d}$ defined by
\begin{align} \label{eq:FS_mod_dist}
{\tilde d}(x_{A}, y_{B}) = \mathbbm{E}[d(X_{B}, y_{B})|X_{A} = x_{A}].
\end{align}

This fact will serve as a stepping stone to our analysis of USRDf for 
a $k$-random sampler. In the Bayesian setting, we consider a modified 
distortion measure $d_{\tau_1}$, $\tau_1 \in \Theta_1$, given by 
\begin{align} \label{eq:modified_distortion}
 d_{\tau_1}(x_{A}, y_{B}) \triangleq \mathbbm{E}[d(X_{B}, y_{B}) | X_{A} = x_{A}, \ \theta_1 = \tau_1 ];
\end{align}
the set of (constrained) pmfs 
\begin{align} \label{eq:const_FS_Bay}
 {\kappa}_{A}^{\cB}(\delta, \tau_1) \triangleq \{ P_{\theta X_{\cM} Y_{B}}: \ \theta, X_{\cM} \MC \theta_1, X_{A} \MC Y_{B}, \ \mathbbm{E}[d_{\tau_1}(X_{A}, Y_{B})| \theta_1 = \tau_1] \leq \delta \},
\end{align}
and the (minimized) conditional mutual information
\begin{align} \label{eq:primitive_k_FS_bayesian}
 \rho_{A}^{\cB}(\delta, \tau_1) \triangleq \underset{ \kappa_{A}^{\cB}(\delta, \tau_1) } \min I(X_{A} \wedge Y_{B} | \theta_1 = \tau_1)
\end{align}
which is akin to \eqref{eq:FS_SRDf} and will play a basal role.  
In the nonBayesian setting, the counterparts of \eqref{eq:const_FS_Bay} 
and \eqref{eq:primitive_k_FS_bayesian} are 
\begin{align} \label{eq:const_FS_nonBay}
 {\kappa}_{A}^{n\cB}(\delta, \tau_1) \triangleq \{ P_{ X_{\cM} Y_{B} | \theta = \tau} 
 = P_{X_{\cM}|\theta = \tau} P_{Y_{B}|X_{A}, \theta_1 = \tau_1}: \ 
 \mathbbm{E}[d (X_{B}, Y_{B})| \theta = \tau] \leq \delta, \ \tau \in \Lambda(\tau_1) \}
\end{align}
and
\begin{align} \label{eq:primitive_k_FS_nonBayesian}
 \rho_{A}^{n\cB}(\delta, \tau_1) \! \triangleq \! 
 \underset{ \kappa_{A}^{n \cB}(\delta, \tau_1) } \min \! I(X_{A} \wedge Y_{B} | \theta_1 = \tau_1).
 \vspace*{-0.3cm}
\end{align}

\noindent {\it Remarks}: (i) The minima in \eqref{eq:primitive_k_FS_bayesian} 
and \eqref{eq:primitive_k_FS_nonBayesian}
exist as those of convex functions over convex, compact sets.

\vspace*{0.1cm}

\noindent (ii) Clearly, the minimum in \eqref{eq:primitive_k_FS_nonBayesian} 
under pmf-wise constraints \eqref{eq:const_FS_nonBay} can be no smaller 
than that in \eqref{eq:primitive_k_FS_bayesian} under pmf-averaged constraints \eqref{eq:const_FS_Bay}.

\vspace*{0.1cm}

\noindent (iii) It is seen in a standard manner that $ \rho_{A}^{\cB}(\delta,\tau_1)$
in \eqref{eq:primitive_k_FS_bayesian} and 
$ \rho_{A}^{n\cB}(\delta,\tau_1)$ in \eqref{eq:primitive_k_FS_nonBayesian} 
are convex and continuous in $\delta.$

\vspace*{0.1cm}

Our first main result states that the USRDf at distortion level $\Delta$ 
for fixed-set sampling in the Bayesian setting is a minmax of quantities 
in \eqref{eq:primitive_k_FS_bayesian}, where the maximum is over ambiguity 
atoms $ \tau_1$ in $\Theta_1$, while the minimum is over distortion
thresholds $\delta = \Delta_{\tau_1}, \ \tau_1 \in \Theta_1$ whose mean 
does not exceed $\Delta$. On the other hand, in the nonBayesian setting,
the USRDf at distortion level $\Delta$ is a maximum over ambiguity atoms 
of quantities in \eqref{eq:primitive_k_FS_nonBayesian} with $\delta = \Delta,$ 
and hence is no smaller than its Bayesian counterpart.

\begin{theorem} \label{prop:k_FS_finite}
 The 
 Bayesian USRDf for fixed $A \subseteq \cM$ is  
 \begin{align}
  R_{A}(\Delta) 
  & \! = \! \underset{ \{ \Delta_{\tau_1} , \ \tau_1 \in \Theta_1 \} 
  \atop \mathbbm{E}[\Delta_{\theta_1}] \leq \Delta } 
  \min \ \underset{\tau_1 \in \Theta_1} \max \ \rho_{A}^{\cB}(\Delta_{\tau_1},\tau_1) 
  \label{eq:Bayesian_kfs}
 \end{align}
for $\ \Delta_{\min} \leq \Delta \leq \Delta_{\max} ,$ where 
\begin{align}
\Delta_{\min} = \mathbbm{E} \Big [ \mathbbm{E} [\underset{y_{B} \in \cY_{B}} 
\min d_{\theta_1}(X_{A}, y_{B})| \theta_1] \Big ] = 
\mathbbm{E}[\underset{y_{B} \in \cY_{B}} \min d_{\theta_1}(X_{A}, y_{B}) ],  
\end{align}
\begin{align}
\Delta_{\max} = \mathbbm{E} \big [\underset{y_{B} \in \cY_{B}} \min 
\mathbbm{E}[d_{\theta_1}(X_{A}, y_{B})|\theta_1] \big].
\end{align}
The nonBayesian USRDf is
\begin{align} \label{eq:nonBayesian_kfs}
  R_{A}(\Delta) 
  & =  \underset{\tau_1 \in \Theta_1} \max \  \rho_{A}^{n \cB}(\Delta , \tau_1), 
  \ \ \  \Delta_{\min} \leq \Delta \leq \Delta_{\max}
 \end{align}
 where 
 \begin{align}
  \Delta_{\min} =  
 \underset{\tau_1 \in \Theta_1} \max \ \underset{ P_{Y_{B}|X_{A}, \theta_1 = \tau_1}
 = P_{Y_{B}|X_{\cM}, \theta = \tau}} \min \ \underset{\tau \in \Lambda(\tau_1)} 
 \max \ \mathbbm{E}[  d (X_{B}, Y_{B}) | \theta = \tau] 
 \end{align}
 and
 \begin{align}
  \Delta_{\max} =  \underset{\tau_1 \in \Theta_1} \max \ \underset{y_{B} \in \cY_{B}} 
  \min \ \underset{\tau \in \Lambda(\tau_1)} \max \  \mathbbm{E}[d (X_{B}, y_{B})|\theta = \tau].
 \end{align}

\end{theorem}

\noindent {\it Remarks}: (i) In fact, the minimizing pmf $P_{Y_B|X_A \theta_1}$ in 
$\Delta_{\min}$ is a conditional point-mass.

\noindent (ii) We note that for a given distortion level 
$\Delta,$ the set  $\{ \Delta_{\tau_1}, \ \tau_1 \in \Theta_1: \ 
\sum \limits_{\tau_1 \in \Theta_1} \mu_{\theta_1}(\tau_1) \Delta_{\tau_1} \leq \Delta \}$ is 
a convex, compact set in $\mathbbm{R}^{|\Theta_1|}$. Next, observing that  
\begin{align}
 \underset{\tau_1 \in \Theta_1} \max \ \rho_{A}^{\cB}(\Delta_{\tau_1}, \tau_1)
\end{align}
is a convex function of $\{ \Delta_{\tau_1}, \ \tau_1 \in \Theta_1 \}$, the 
minimum in \eqref{eq:Bayesian_kfs} exists as that of a
convex function over a convex, compact set.

\vspace*{0.1cm}

\noindent (iii) The minimizing $\{ \Delta_{\tau_1}^{*}, \ 
\tau_1 \in \Theta_1 \}$ in \eqref{eq:Bayesian_kfs}
is characterized by the following special property: For a given 
$\Delta_{\min} \leq \Delta \leq \Delta_{\max},$
for each $\tau_1 \in \Theta_1,$ either
\begin{align} \label{eq:Bay_kfs_rem3}
 \rho_{A}^{\cB}(\Delta_{\tau_1}^{*}, \tau_1) \equiv 
 \underset{ {\tilde \tau}_1 \in \Theta_1} \max \ 
 \rho_{A}^{\cB}(\Delta_{{\tilde \tau}_1}^{*}, {\tilde \tau}_1)
\end{align}
where the right-side does not depend on $\tau_1$, or
\begin{align}
 \Delta_{\tau_1}^{*} = \mathbbm{E}[  \underset{y_{B} \in \cYB} 
 \min d_{\tau_1}(X_{A}, y_{B})|\theta_1 = \tau_1 ].
\end{align}
By a standard argument in convex optimization, if 
$\{ \Delta_{\tau_1}^{*}, \ \tau_1 \in \Theta_1 \}$
does not satisfy the property above, then a small perturbation 
decreases the maximum in \eqref{eq:Bay_kfs_rem3}
leading to a contradiction.

\vspace*{0.1cm}

\noindent (iv) The $\Delta_{\min}$ and $\Delta_{\max}$ for the Bayesian and the 
nonBayesian settings can be different.

\vspace*{0.2cm}

\begin{example}

For the probability of error distortion measure 
\begin{align}
 d(x_B, y_B) = \mathbbm{1}(x_B \neq y_B) = 1 - \prod \limits_{i \in B} \mathbbm{1}(x_i = y_i),
 \quad x_B, y_B \in \cX_B = \cY_B
\end{align}
the Bayesian USRDf for fixed-set sampling with $A \subseteq B$ in \eqref{eq:Bayesian_kfs} 
simplifies with \eqref{eq:primitive_k_FS_bayesian} becoming
\begin{align} \label{ex1:Bay_prim}
 \rho_{A}^{\cB}(\Delta_{\tau_1}, \tau_1) = 
 \underset{ \mathbbm{E}[ \alpha_{\tau_1}(X_A) 
 \mathbbm{1}(X_A \neq Y_A) | \theta_1 = \tau_1 ]
 \leq \Delta_{\tau_1} - (1 - \mathbbm{E}[\alpha_{\tau_1}(X_{A}) |
 \theta_1 = \tau_1]) } \min I(X_{A} \wedge Y_{A} | \theta_1 = \tau_1)
\end{align}
where 
\begin{align} \label{ex1:alpha_def}
 \alpha_{\tau_1}(x_{A}) = \underset{ {\tilde x} \in \cX_B} 
 \max \ P_{X_B | X_{A} \theta_1} ( {\tilde x}|x_A, \tau_1)
\end{align}
is the {\it maximum a posteriori} (MAP) estimate of $X_B$ 
on the basis of $X_A = x_A$ under pmf
$P_{X_{\cM}|\theta_1 = \tau_1}$.

\vspace*{0.1cm}

\noindent The proof of \eqref{ex1:Bay_prim}, \eqref{ex1:alpha_def} 
is along the lines of that of
(\cite{BodNar17}, Proposition 1) under the pmf 
$P_{X_{\cM}|\theta_1 = \tau_1}$ (rather than $P_{X_{\cM}}$
as in \cite{BodNar17}), and so is not repeated here. Furthermore,
\begin{align}
 \Delta_{\min} =  1 - \mathbbm{E}[\alpha_{\theta_1}(X_{A}) ] \quad \text{ and } \quad 
 \Delta_{\max} =  1 - \mathbbm{E} \big [  \underset{x_{B} \in \cX_B} \max 
 \ P_{X_{B}|\theta_1}(x_B | \theta_1) \big ].
\end{align}

The form of the Bayesian USRDf in \eqref{ex1:Bay_prim}  
suggests a simple achievability
scheme comprising two steps. Using a {\it maximum a posteriori} (MAP) 
or {\it maximum likelihood} (ML)
estimate ${\widehat \tau}_1$ of $\theta_1$ on the basis of 
$X_A^n = x_A^n$, the first step entails a lossy 
reconstruction of $x_{A}^n$ by its codeword $y_{A}^n$, 
under pmf $P_{X_{\cM}|\theta_1 = {\widehat \tau}_1}$
and for a modified distortion measure
\begin{align}
 {\tilde d}_{{\widehat \tau}_1}(x_A, y_A) \triangleq 
 \alpha_{{\widehat \tau}_1}(x_A) \mathbbm{1}(x_A \neq y_A)
\end{align}
with a corresponding reduced threshold
\begin{align}
 \Delta_{{\widehat \tau}_1} - (1 - \mathbbm{E}[\alpha_{ {\widehat \tau}_1}(X_{A}) | \theta_1 = {\widehat \tau}_1]).
\end{align}
This is followed by a second step of reconstructing $x_{B}^n$ from the output $y_A^n$
of the previous step as a MAP estimate
\begin{align}
 y_B^n = \underset{y^n \in \cY_{B}^n} {\arg \max} \ P_{X_B|X_A \theta_1} (y^n| y_A^n, {\widehat \tau}_1);
\end{align}
the corresponding probability of estimation error coincides with the 
mentioned reduction $1 - \mathbbm{E}[ \alpha_{{\widehat \tau}_1}(X_{A})|\theta_1 = {\widehat \tau}_1]$
in the threshold.

\vspace*{0.1cm}

In the nonBayesian setting, the USRDf in \eqref{eq:nonBayesian_kfs},
\eqref{eq:primitive_k_FS_nonBayesian} simplifies with 
\begin{align} \label{ex1:nonBayesian}
 \rho_{A}^{n \cB}(\Delta, \tau_1) = \underset{ P_{Y_{A}|X_A, \theta_1 = \tau_1} P_{Y_{B \setminus A}|Y_A, \theta_1 = \tau_1}
  = P_{Y_B | X_{\cM}, \theta = \tau} \atop 
  \mathbbm{E}[  \mathbbm{1}(X_B \neq Y_B)  | \theta = \tau ] 
\leq \Delta ,
\quad \tau \in \Lambda(\tau_1) }  \min \ I(X_{A} \wedge Y_A | \theta_1 = \tau_1),
\end{align}
for $\Delta_{\min} \leq \Delta \leq \Delta_{\max}$, where
\begin{align}
 \Delta_{\min} = \underset{\tau_1 \in \Theta_1} \max \ \underset{ P_{Y_B|X_A, \theta_1= \tau_1} } \min \ \underset{\tau \in \Lambda(\tau_1)} \max \big (1 - P(X_B = Y_B | \theta = \tau) \big )
\end{align}
and
\begin{align}
  \Delta_{\max} = \underset{\tau_1 \in \Theta_1} \max \underset{y_B \in \cY_B} \min
 \underset{ \tau \in \Lambda(\tau_1)} \max \big ( 1 - P_{X_B|\theta}(y_B|\tau) \big ).
\end{align}
This leads to the following achievability scheme. With ${\widehat \tau}_1$ as the ML
estimate of $\theta_1$ formed from $X_A^n = x_A^n$, first $x_A^n$ is reconstructed
as $y_A^n$ according to $P_{Y_A|X_A , \theta_1 = {\widehat \tau}_1}$ resulting from the 
minimization in \eqref{ex1:nonBayesian}.
This is followed by the reconstruction of $x_B^n$ from $y_A^n$ by means of the 
estimate
\begin{align}
 y_B^n = \underset{y^n \in \cY_B^n} {\arg \max} \ P_{Y_B | Y_A \theta_1}(y^n | y_A^n , {\widehat \tau}_1)
\end{align}
under pmf $P_{Y_B | Y_A \theta_1}$ which, too, is obtained from the minimization 
in \eqref{ex1:nonBayesian}.

\qed

\end{example}

\begin{figure}[h]
\centering
  \includegraphics[width=6.2cm,height=3.1cm]{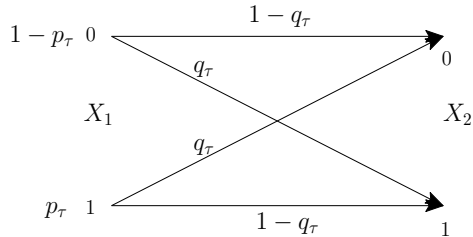}  
  \caption{Virtual BSC ($q$)}
  \label{fig:Virtual_BSC}
\end{figure}

\begin{example}
 
Let $\cM  = \{1,2\}$ and  ${\cX}_{1}= \cX_2 = \{0,1\},$ consider a 
DMMS with  $P_{X_{1} X_{2}|\theta = \tau} $ represented  
by a virtual binary symmetric channel (BSC) shown in Figure \ref{fig:Virtual_BSC},
where $p_{\tau}, q_{\tau} \leq 0.5, \ \tau \in \Theta$, 
where $\Theta$ is a given finite set.
For $A = \{1\}, \ B = \{1,2\}$, and the probability of error distortion measure
of Example 1, the Bayesian USRDf reduces to
\vspace*{-0.1cm}
\begin{align} \label{eq:ex1_BayUSRDf}
 R_{\{1\}}(\Delta) \! = \! \underset{ \{\Delta_{\tau_1}, \ \tau_1 \in \Theta_1 \} \atop \mathbbm{E}[\Delta_{\theta_1}] \leq \Delta } \min  \underset{\tau_1 \in \Theta_1} \max  \Big ( h(p_{\tau_1}) -
  h \Big ( \frac{\Delta_{\tau_1} - q_{\tau_1}}{1 - q_{\tau_1}} \Big )
 \Big )  \! ,
\end{align}
for $\Delta_{\min} \leq \Delta \leq \Delta_{\max}$, where
\begin{align}
 \Delta_{\min} = \mathbbm{E}[q_{\theta_1}], \quad  \Delta_{\max} = \mathbbm{E}[p_{\theta_1} + q_{\theta_1} - p_{\theta_1} q_{\theta_1}];
\end{align}
and $q_{\tau_1} = P_{X_{2}|X_{1} \theta_1 }(0|1,\tau_1), \ \tau_1 \in \Theta_1$; 
and the nonBayesian USRDf is
\begin{align}
 R_{\{1\}}(\Delta) = \underset{\tau_1 \in \Theta_1} \max \Big ( h(p_{\tau_1}) -
 \underset{\tau \in \Lambda(\tau_1)} \min h \Big ( \frac{\Delta - q_{\tau}}{1 - q_{\tau}} \Big )  \Big )
\end{align}
with
\begin{align}
 \Delta_{\min} =  \underset{ \tau \in \Theta} \max \ q_{\tau}  \quad \text{ and } \quad 
 \Delta_{\max} =  \underset{\tau \in \Theta} \max \ ( p_{\tau} + q_{\tau} - p_{\tau} q_{\tau}).
\end{align}

\qed 

\end{example}

\vspace*{-0.5cm}

\vspace*{0.4cm}

\begin{example}

This example, albeit concocted, shows that for fixed-set sampling with $A$ and recovery set $B$,
a choice of $A$ outside $B$ can be best.
Let $\cM = \{1,2,3\}, \ B = \{1,2\}$ and ${\cX}_{i} = \cY_{j} = \{ 0,1 \}, \ i = 1,2,3; j = 1,2.$ 
Consider a DMMS with $P_{X_{1} X_{2} | \theta = \tau}$ as in
Figure \ref{fig:Virtual_BSC} and $X_{3} = X_{1} \oplus X_{2}$
where $\oplus$ denotes addition modulo 2.
Here, $p_{\tau} = 0.5, \ q_{\tau} \leq 0.5, \ \tau \in \Theta$,
with the $q_{\tau}$s$, \ \tau \in \Theta$, being distinct.
For distortion measure
 $d(x_{B},y_{B}) \triangleq \mathbbm{1}\left( (x_{1} \oplus x_{2}) \neq (y_{1} \oplus y_{2}) \right)$, 
the Bayesian USRDf for fixed-set sampling is
\begin{align}
 R_{ \{1\}}(\Delta) = 
 h(0.5) - h \Big ( \dfrac{\Delta - {\tilde q}}{1 - 2 {\tilde q}}  \Big ),
 \ \ \ {\tilde q} \leq \Delta \leq 0.5,
\end{align}
where ${\tilde q} = \sum \limits_{\tau \in \Theta} \mu_{\theta}(\tau) q_{\tau}$.
Since $P_{X_{1}|\theta = \tau}$ is the same for all $\tau \in \Theta,$
note that $|\Theta_1 | = 1.$
The nonBayesian USRDf is
\begin{align}
 R_{ \{1\}}(\Delta) = 
 h(0.5) - \underset{\tau \in \Theta} \min \ h \Big ( \dfrac{\Delta - q_{\tau}}{1 - 2 q_{\tau}}  \Big ),
 \quad \underset{\tau \in \Theta} \max \ q_{\tau} \leq \Delta \leq 0.5 .
\end{align}
Also, $R_{\{1\}}(\Delta) = R_{\{2\}}(\Delta)$.
 For sampling set $A = \{3\}$, $\Theta_1 = \Theta$ and the Bayesian USRDf is  
 \begin{align*}
 R_{ \{3\}}(\Delta) = \underset{ \{\Delta_{\tau}, \ \tau \in \Theta\}: \ 
\mathbbm{E}[\Delta_{\theta}] \leq \Delta } \min  \ \underset{\tau \in \Theta} \max 
 \ h(q_{\tau}) - h \left (  \Delta_{\tau} \right ), \quad 0 \leq \Delta \leq {\tilde q}, 
\end{align*}
and the nonBayesian USRDf is
\begin{align*}
 R_{ \{3\}}(\Delta) = \underset{\tau \in \Theta} \max \ h(q_{\tau}) - h \left (  \Delta \right ), \ \
\ 0 \leq \Delta \leq \underset{\tau \in \Theta} \max \ q_{\tau}.
 \end{align*}
 Clearly, $R_{ \{3 \}}(\Delta) \leq R_{ \{1 \}}(\Delta)$,
 with the inequality being strict for suitable values of $\Delta.$
 \qed
\end{example}

\vspace*{0.2cm}

Turning to a $k$-IRS in \eqref{eq:k_IRS_def}, the freedom now given to the sampler to rove over
all $k$-sized subsets in $\cAk$ engenders a partition $\Theta_2$ of $\Theta_1$ (and hence
a finer partition of $\Theta$) with smaller ambiguity atoms. Let $A_{1}, \ldots, A_{|\cAk|},$
where $|\cAk| = {m \choose k}$, be any fixed ordering of $\cAk.$
Let $\Theta_2$ be a partition of $\Theta$ consisting of ambiguity atoms,
with each atom formed by $\tau$s with {\it identical (ordered) collections of marginal pmfs}
$\left ( P_{X_{A_{i}}|\theta = \tau}, \ i=1, \ldots, |\cAk| \right ).$

\vspace*{0.2cm}

Clearly, $\Theta_2$ is a refinement
of $\Theta_1$ (for any $A_{i}$). Indexing the elements of $\Theta_2$ by $\tau_2,$
let $\theta_2$ be a $\Theta_2$-valued
rv with pmf $\mu_{\theta_2}$ derived from $\mu_{\theta}$.
For each $\tau_2$ in $ \Theta_2,$ let $\Lambda(\tau_2)$ be the collection of $\tau$s in the atom indexed by $\tau_2$. In analogy with \eqref{eq:primitive_k_FS_bayesian} and \eqref{eq:primitive_k_FS_nonBayesian},
we define counterparts in the Bayesian and nonBayesian settings as 
\begin{align} \label{prim:IRS_bayesian}
& \rho_{\imath}^{\cB}(\delta,P_{S}, \tau_2)  
  \! \triangleq \! \underset{ \kappa_{\imath}^{\cB}(\delta,P_{S}, \tau_2)  } 
  \min \ I(X_{S} \wedge Y_{B} |S, \theta_2 = \tau_2); \\
 & \rho_{\imath}^{n\cB}(\delta,P_{S}, \tau_2)  
  \! \triangleq \!  \underset{ \kappa_{\imath}^{n\cB}(\delta, P_{S},\tau_2) } \min \ I(X_{S} \wedge Y_{B} |S, \theta_2 = \tau_2),  \label{prim:IRS_nonBayesian} 
 \vspace*{-1cm}
 \vspace*{-0.1cm}
\end{align}
where $d_{\tau_2}$ is defined  as in \eqref{eq:modified_distortion} with $\theta_2 = \tau_2$ replacing $\theta_1  =\tau_1,$
and
\begin{align} 
 {\kappa}_{\imath}^{\cB}(\delta,P_{S}, \tau_2) \triangleq \big \{ P_{\theta X_{\cM} S Y_{B}} = \mu_{\theta} P_{X_{\cM}|\theta } P_{S}    P_{Y_{B}|S X_{S}  \theta_2 }   :   \sum \limits_{A \in \cAk} P_{S}(A) \mathbbm{E}[d_{\tau_2}(X_{A}, Y_{B})| S = A,\theta_2 = \tau_2] \leq \delta \big \}, \label{eq:const_IRS_Bay}  
 \end{align}
 \vspace*{-0.7cm}
 \begin{align}  
 {\kappa}_{\imath}^{n\cB}(\delta, P_{S},\tau_2) \triangleq  \big \{P_{X_{\cM} S Y_{B}|\theta = \tau} = P_{X_{\cM}|\theta = \tau} P_{S}  & P_{Y_{B}|S X_{S}, \theta_2 = \tau_2}:  \\ & \sum \limits_{A \in \cAk} P_{S}(A) \mathbbm{E}[d (X_{B}, Y_{B})| S = A,\theta = \tau] \leq \delta , \ \tau \in \Lambda(\tau_2) \big \}.  \label{eq:const_IRS_nonBay}
\end{align}

\begin{theorem} \label{th:k_IRS_finite}
 The Bayesian USRDf for a $k$-IRS is
 \begin{align}
 \label{eq:USRDf_IRS_Bayesian}
  R_{\imath}(\Delta)  =  \underset{ P_{S}, \ \{ \Delta_{\tau_2}, \ \tau_2 \in \Theta_2 \}  \atop 
  \mathbbm{E}[\Delta_{\theta_2}] \leq \Delta }
  \min \ \underset{\tau_2 \in \Theta_2} \max \ \rho_{\imath}^{\cB} (\Delta_{\tau_2},P_{S},\tau_2), \ \ \ \ \Delta_{\min} \leq \Delta \leq \Delta_{\max},
 \end{align} 
 where
 \begin{align}
  \Delta_{\min} = \underset{ A  \in \cAk } \min \ \mathbbm{E} \big[ \mathbbm{E}[ \underset{y_{B} \in \cY_{B}} \min d_{\theta_2}(X_{A }, y_{B}) |  \theta_2  ] \big ] 
\text{ and }
  \Delta_{\max} = \underset{A \in \cAk} \min \  \mathbbm{E}\big[ \underset{y_{B} \in \cY_{B}} \min \mathbbm{E}[d_{\theta_2}(X_{A}, y_{B})|\theta_2 ] \big].
 \end{align}
The nonBayesian USRDf is 
\begin{align} \label{eq:USRDf_IRS_nonBaye_fin}
  R_{\imath}(\Delta)  =  \underset{ P_{S} } \min \ \underset{\tau_2 \in \Theta_2} \max  \
   \rho_{\imath}^{n\cB}(\Delta,P_{S},\tau_2), \ \ \ \ \Delta_{\min} \leq \Delta \leq \Delta_{\max},
 \end{align}
 for 
 \begin{align}
  \Delta_{\min} = \underset{P_{S}} \min \ \underset{\tau_2 \in \Theta_2} \max  \sum \limits_{A \in \cAk} P_{S}(A) \underset{P_{Y_{B}| S X_{S}, \theta_2 = \tau_2} = P_{Y_{B} | S X_{\cM}, \theta = \tau}} \min \ \underset{\tau \in \Lambda(\tau_2)} \max \mathbbm{E}[  d (X_{B}, Y_{B}) | S = A, \theta = \tau] 
 \end{align}
and 
 \begin{align}
  \Delta_{\max} =  \underset{\tau_2 \in \Theta_2} \max \ \underset{y_{B} \in \cY_{B}} \min \ \underset{\tau \in \Lambda(\tau_2)} \max \  \mathbbm{E}[d (X_{B}, y_{B})|\theta = \tau].
 \end{align}

\end{theorem}

\begin{corollary}
 The USRDfs in the Bayesian and nonBayesian settings remain unchanged upon
 a restriction to $n$-length block codes $(f, \varphi)$ with uninformed
 decoder, i.e., with $\varphi = \varphi(f(S^n, X_{S}^{n}))$.
\end{corollary}

\noindent {\it Remark}: (i) For a $k$-IRS we restrict ourselves to the 
interesting case  of $k < |B|,$ for otherwise it would suffice to 
choose $S_t = B, \ t = 1 ,\ldots,n.$

\noindent (ii)  Akin to a $k$-FS, the optimizing $P_{S}, \ \{ \Delta_{\tau_2}^{*}, \ \tau_2 \in \Theta_2 \}$ in \eqref{eq:USRDf_IRS_Bayesian}
has the following special property: For a given $\Delta_{\min} \leq \Delta \leq \Delta_{\max},$
for each $\tau_2 \in \Theta_2,$ either
\begin{align} \label{eq:Bay_IRS_rem3}
 \rho_{\imath}^{\cB}(\Delta_{\tau_2}^{*}, P_{S}, \tau_2) = \underset{ {\tilde \tau}_2 \in \Theta_2} \max \ \rho_{\imath}^{\cB}(\Delta_{{\tilde \tau}_2}^{*}, P_{S},  {\tilde \tau}_2)
\end{align}
or
\begin{align}
 \Delta_{\tau_2}^{*} = \sum \limits_{A \in \cAk} P_{S}(A)\mathbbm{E}[  \underset{y_{B} \in \cYB} \min d_{\tau_2}(X_{A}, y_{B})|\theta_2 = \tau_2 ].
\end{align}

\noindent (iii) In general, a $k$-IRS will outperform a $k$-FS in two ways. First, the former enables
a better approximation of $\theta$ in the form of $\theta_2$ whereas the latter estimates
$\theta_1 = \theta_1(\theta_2)$. Second, random sampling enables a ``time-sharing'' over various fixed-set 
samplers, that can outperform strictly the best fixed-set choice. Both these advantages of
a $k$-IRS over fixed-set sampling are illustrated in Examples 4 and 5.

\vspace*{0.2cm}

\begin{example}
 This example illustrates that a $k$-IRS can perform  strictly better than the best
 $k$-FS.
 For $\cM = B  = \{1,2\}$, and $\cX_{i} = \cY_{i} = \{0,1\}, \ i = 1,2,$ 
 consider a DMMS with $P_{X_{1} X_{2}|\theta = \tau} =  P_{X_{1} |\theta = \tau} P_{X_{2}|\theta = \tau} $
where
\begin{align*}
 P_{X_{1}|\theta }( 0|\tau) = 1 - p_{\tau}, \ \ P_{X_{2}|\theta }( 0|\tau) = 1 - q_{\tau}, \ 
 \tau \in \Theta,
\end{align*}
and  $0 < p_{\tau}, \ q_{\tau} < 0.5$.
Under the  distortion measure
 $d(x_{B}, y_{B}) = \mathbbm{1}(x_{1} \neq y_{1} ) + \mathbbm{1}(x_{2} \neq y_{2} )$,
for a $k$-FS, with $k=1$, the Bayesian USRDf for sampling set $A = \{1\}$ is
\begin{align*}
\quad  R_{ \{1\}}(\Delta) \! =  \! \underset{  \{\Delta_{\tau_1}, \ 
\tau_1 \in \Theta_1 \} \atop \mathbbm{E}[\Delta_{\theta_1}] \leq \Delta } 
\min  \underset{ \tau_1 \in \Theta_1} 
  \max \ \Big ( \! h(p_{\tau_1}) \! - \! h \big  ( \Delta_{\tau_1} \! - 
  \! q_{\tau_1} \big ) \! \Big ), \quad \mathbbm{E}[q_{\theta }] \leq \Delta 
  \leq \mathbbm{E}[p_{\theta } + q_{\theta }]
\end{align*}
where $q_{\tau_1} = \mathbbm{E}[q_{\theta} | \theta_1 = \tau_1]$, and the nonBayesian USRDf is
\begin{align}
\quad  R_{ \{1\}}(\Delta) = \underset{ \tau_1 \in \Theta_1} \max \left ( h(p_{\tau_1}) - 
  \underset{\tau \in \Lambda(\tau_1)} \min h \big  ( \Delta - q_{\tau} \big ) \right ) 
  \quad  \underset{\tau \in \Theta} \max \ q_{\tau} \leq \Delta \leq 
  \underset{\tau \in \Theta} \max \ (p_{\tau} + q_{\tau}   ).
\end{align}

Turning to a $k$-IRS with $k=1$, clearly, $\Theta_2 = \Theta$.  
For a $k$-IRS the Bayesian USRDf is
\begin{align} 
\quad R_{\imath}(\Delta) = \underset{P_{S}, \  \{\Delta_{\tau}, \ \tau \in \Theta \}
\atop \mathbbm{E}[\Delta_{\theta}] \leq \Delta} \min \underset{ \tau \in \Theta} 
 \max \! \underset{ \Delta_{1 \tau}, \ \Delta_{2 \tau} \atop 
  P_{S}(\{1\}) \Delta_{1 \tau} +  P_{S}(\{2\}) \Delta_{2 \tau} 
  \leq \Delta_{\tau} } \min I,
     \quad \min \{ \mathbbm{E}[ p_{\theta}], \mathbbm{E}[q_{\theta}] \}  
     \leq \Delta \leq \mathbbm{E}[p_{\theta} + q_{\theta}]
\end{align}
and the nonBayesian USRDf  is
\begin{align} \label{ex3:nonBayesian_IRS}
\quad R_{ \imath}(\Delta) & = \underset{P_{S} } \min \ \underset{ \tau \in \Theta} 
 \max  \! \underset{ \Delta_{1 \tau}, \ \Delta_{2 \tau} \atop 
  P_{S}(\{1\}) \Delta_{1 \tau} +  P_{S}(\{2\}) \Delta_{2 \tau} \leq \Delta  } \min  \ I,
  \quad \quad  \underset{ 0 \leq \alpha \leq 1} \min \underset{\tau \in \Theta} \max (\alpha p_{\tau} + (1-\alpha) q_{\tau}) \leq \Delta \leq \underset{\tau \in \Theta} \max \ (p_{\tau} + q_{\tau}   ) \quad 
\end{align} 
 where $I$ equals
 \vspace*{-0.1cm}
 \begin{align*}
  & \! P_{S}(\{1\}) \big ( h(p_{\tau}) - h \big  ( \Delta_{1 \tau} - q_{\tau} \big ) \big )  
     +   P_{S}(\{2\}) \big (h(q_{\tau}) -  h \big  ( \Delta_{2 \tau} -  p_{\tau} \big ) \big ).  
 \end{align*}
An analytical comparison of the USRDfs shows the strict superiority of the $k$-IRS over the $k$-FS, 
as seen -- for instance -- by the lower values of $\Delta_{\min}$ for the former.
 \qed
\end{example}

\vspace*{0.1cm}

\begin{example}
In Example 4, assume that
 \begin{align}
  p_{\tau} \geq q_{\tau} , \quad \tau \in \Theta.
 \end{align}
For a $k$-FS with $k = 1$, the nonBayesian USRDf is
\begin{align} \label{ex3:k_FS_USRDf}
 R_{ \{1\}}(\Delta) = \underset{ \tau_1 \in \Theta_1} \max \left ( h(p_{\tau_1}) - 
  \underset{\tau \in \Lambda(\tau_1)} \min h \big  ( \Delta - q_{\tau} \big ) \right ), \ \ 
  R_{ \{2\}}(\Delta) = \underset{ \tau_1 \in \Theta_1} \max \left ( h(q_{\tau_1}) - 
  \underset{\tau \in \Lambda(\tau_1)} \min h \big  ( \Delta - p_{\tau} \big ) \right ). \quad
\end{align}
Now, observe that for each $\tau \in \Theta $ 
\begin{align}
h(p_{\tau}) - h( \delta - q_{\tau}) \leq h(q_{\tau}) - h( \delta - p_{\tau})
\end{align}
holds for  $ p_{\tau} \leq \delta \leq p_{\tau} + q_{\tau} $.
Thus, for a $k$-IRS with $k=1$, the nonBayesian USRDf   in \eqref{ex3:nonBayesian_IRS}
simplifies to
\begin{align}
 R_{\imath}(\Delta) = \underset{\tau \in \Theta} \max \ h(p_{\tau}) - h(\Delta - q_{\tau})
\end{align}
which is strictly smaller than the USRDf for the better $k$-FS in \eqref{ex3:k_FS_USRDf}.
The superior performance of the $k$-IRS is enabled by its ability to estimate simultaneously
both $P_{X_1 |\theta}$ and $P_{X_2 | \theta}$ (and thereby $P_{X_1 X_2 |\theta}$);
a $k$-FS can estimate only one of $P_{X_1 | \theta}$ or $P_{X_2 | \theta}$.
\qed
\end{example}

Lastly, for a $k$-MRS in \eqref{eq:k_MRS_def}, the ability of the sampler to depend instantaneously
on the current realization
of the DMMS enables an encoder with access to the sampler output to distinguish
among all the pmfs in $\cP.$ Accordingly, for a $k$-MRS, $\Theta$ itself serves as the 
counterpart of the partitions $\Theta_1$ (for a $k$-FS) and $\Theta_2$ for a $k$-IRS.
For a rv $U$ with fixed pmf $P_{U}$ on some finite set $\cU,$ and for fixed $P_{S|X_{\cM} U}$,
we define the counterparts
of \eqref{prim:IRS_bayesian} and \eqref{prim:IRS_nonBayesian} as
\begin{align} \label{prim:MRS_Bayesian}
 \rho_{m}^{\cB} (\delta, P_{U}, P_{S|X_{\cM} U}, \tau) \triangleq \underset{ \kappa_{m}^{\cB}(\delta, P_{U}, P_{S|X_{\cM} U}, \tau) } \min I(X_{S} \wedge Y_{B} | S, U, \theta = \tau ),
\end{align}
and
\begin{align} \label{prim:MRS_nonBayesian}
 \rho_{m}^{n\cB} (\delta, P_{U}, P_{S|X_{\cM} U}, \tau) \triangleq \underset{ \kappa_{m}^{n\cB}(\delta, P_{U}, P_{S|X_{\cM} U}, \tau) } \min I(X_{S} \wedge Y_{B} | S, U, \theta = \tau ),
\end{align}
where the minimization in \eqref{prim:MRS_Bayesian} and \eqref{prim:MRS_nonBayesian}, in effect,
is with respect to $P_{Y_{B}|S X_{S} U \theta}$ and the sets of (constrained) pmfs are
\begin{align}
 \kappa_{m}^{\cB}(\delta, P_{U}, P_{S|X_{\cM} U}, \tau) \triangleq \{  P_{ \theta U X_{\cM} S Y_{B}  } = \mu_{\theta} P_{U} P_{X_{\cM}|\theta } P_{S|X_{\cM} U} P_{Y_{B}|S X_{S} U \theta }: \   
 \mathbbm{E}[d (X_{B}, Y_{B}) | \theta = \tau ]  \leq \delta  \},
\end{align}
and
\begin{align}
 \kappa_{m}^{n\cB}(\delta, P_{U}, P_{S|X_{\cM} U}, \tau) \triangleq \{  P_{ U X_{\cM} S Y_{B} |\theta = \tau} = P_{U} P_{X_{\cM}|\theta = \tau} P_{S|X_{\cM} U} P_{Y_{B}|S X_{S} U, \theta = \tau }: \   
 \mathbbm{E}[d (X_{B}, Y_{B}) |\theta = \tau ]  \leq \delta  \}.
\end{align}
Here, $U$ plays the role of a ``time-sharing'' rv, as will be seen below.

\vspace*{0.1cm}

\begin{theorem} \label{th:k_MRS_finite}
 For a $k$-MRS, the Bayesian USRDf is
 \begin{align} \label{eq:USRDf-MRS-Bayesian-finite}
  R_{m} (\Delta) = \underset{  P_{U}, P_{S|X_{\cM} U}, \{ \Delta_{\tau},  \ \tau \in \Theta \} \atop 
  \mathbbm{E}[\Delta_{\theta}] \leq \Delta}
  \min \ \underset{\tau \in \Theta} \max \ \rho_{m}^{\cB} ( \Delta_{\tau}, P_{U}, P_{S|X_{\cM} U}, \tau), \ \ \ \Delta_{\min} \leq \Delta \leq \Delta_{\max}
 \end{align}
 where 
 \begin{align} \label{eq:MRS_fin_Dmin}
  \Delta_{\min} = \underset{P_{S|X_{\cM}}} \min \mathbbm{E} \Big [ \underset{y_{B} \in \cY_{B} } \min \mathbbm{E} \big [  d  \big (  X_{B}, y_{B} \big ) \big | S, X_{S}, \theta \big ] \Big ]
\text{ and  }
  \Delta_{\max} = \underset{P_{S|X_{\cM}}} \min \mathbbm{E} \Big [  \underset{y_{B} \in \cY_{B} } \min  \mathbbm{E} \big [ d  \big (  X_{B}, y_{B} \big ) \big | S,   \theta \big ] \Big ].
 \end{align}
The nonBayesian USRDf is 
 \begin{align} \label{eq:USRDf-MRS-nonBayesian-finite}
  R_{m} (\Delta) = \underset{  P_{U}, P_{S|X_{\cM} U} }
  \min  \underset{\tau \in \Theta} \max \ \rho_{m}^{n\cB} (\Delta, P_{U}, P_{S|X_{\cM} U}, \tau), \quad \Delta_{\min}
  \leq \Delta \leq \Delta_{\max}, 
 \end{align}
 where 
  \begin{align} \label{eq:MRS_fin_Dmin_nB}
  \Delta_{\min} 
     & = \underset{P_{S|X_{\cM}}} \min \ \underset{\tau \in \Theta} \max \  \mathbbm{E} \big [ \underset{y_{B} \in \cY_{B} } \min \mathbbm{E} \big [   d (X_{B}, y_{B}) |S  , X_{S} , \theta = \tau \big ] \big | \theta = \tau \big ]
 \end{align}
 and 
 \begin{align} \label{eq:MRS_fin_Dmax_nB}
  \Delta_{\max} = \underset{P_{S|X_{\cM}}} \min \ \underset{\tau \in \Theta} \max \ \sum \limits_{A_{i} \in \cAk} P_{S|\theta}(A_{i}|\tau)  \underset{y_{B} \in \cY_{B} } \min \mathbbm{E} \big [ d (X_{B}, y_{B}) |S = A_{i}, \theta = \tau \big ]  .
 \end{align}
 It suffices to take $|\cU| \leq 2|\Theta| + 1.$
\end{theorem}
 
In \eqref{eq:MRS_fin_Dmin} and \eqref{eq:MRS_fin_Dmin_nB}, \eqref{eq:MRS_fin_Dmax_nB}, 
it is readily seen that conditionally deterministic 
samplers (defined below) attain the minima in $\Delta_{\min}$ and
$\Delta_{\max}$.
In fact, such samplers will be seen to be optimal for every  
$\Delta_{\min} \leq \Delta \leq \Delta_{\max}$.

\vspace*{0.2cm}

For a mapping $w : {\cal X}_{\cM} \times {\cal U} \rightarrow 
 {\cal A}_{k}$,  a {\it deterministic sampler} is specified in terms of a conditional point-mass pmf
 \begin{equation}
 \label{eq:point_mass_sampler}
 P_{S|X_{\cM} U}(s|x_{\cM}, u) = \delta_{  w(x_{\cM}, u) }(s)  \triangleq 
 \begin{cases}
       1, \  & s = w(x_{\cM}, u)  \\
       0, \  & \text{otherwise}, \hspace*{0.5cm} \  
       (x_{\cM},u) \in {\cal X}_{\cM} \times {\cal U}, \ s \in \cAk. 
 \end{cases}
 \end{equation}

\noindent Theorem \ref{th:k_MRS_finite}  is equivalent to
 
\noeqref{eq:point_mass_sampler}

\begin{proposition} \label{prop:MRS_fin_opt_samp}
For a $k$-MRS, the Bayesian USRDf is
 \begin{align}  \label{eq:USRDf-MRS-Bayesian-finite_alt}
  R_{m} (\Delta) = \underset{  P_{U}, \ \delta_{w }, \{ \Delta_{\tau},  \ \tau \in \Theta \} \atop 
  \mathbbm{E}[\Delta_{\theta}] \leq \Delta}
  \min \ \underset{\tau \in \Theta} \max \ \rho_{m}^{\cB} ( \Delta_{\tau}, P_{U}, \delta_{w }, \tau), \ \ \ \Delta_{\min} \leq \Delta \leq \Delta_{\max}
 \end{align}
with $\Delta_{\min}$ and $\Delta_{\max}$ as in \eqref{eq:MRS_fin_Dmin},
and the nonBayesian USRDf is
\begin{align}  \label{eq:USRDf-MRS-nonBayesian-finite_alt}
  R_{m} (\Delta) = \underset{  P_{U}, \ \delta_{w } }
  \min \ \underset{\tau \in \Theta} \max \ \rho_{m}^{n\cB} ( \Delta , P_{U}, \delta_{w }, \tau), \ \ \ \Delta_{\min} \leq \Delta \leq \Delta_{\max}
 \end{align}
with $\Delta_{\min}$ and $\Delta_{\max}$ as in \eqref{eq:MRS_fin_Dmin_nB}
and \eqref{eq:MRS_fin_Dmax_nB}, respectively.
It suffices if $|\cU| \leq 2|\Theta | + 1.$

\end{proposition}

\noindent {\it Proof}: See Appendix \ref{app:prop1}.

\vspace*{0.2cm}

\noindent The achievability proof of Theorem \ref{th:k_MRS_finite}, by dint of
Proposition \ref{prop:MRS_fin_opt_samp}, will use a deterministic sampler 
based on the minimizing $w$ from \eqref{eq:USRDf-MRS-Bayesian-finite_alt} or
\eqref{eq:USRDf-MRS-nonBayesian-finite_alt}.

\vspace*{0.2cm}

\begin{example}
 This example compares the USRDfs for a $k$-MRS and a $k$-IRS and is an adaptation of Example 2 above 
 (and also of (\cite{BodNar17}, Example 2)). Consider Example 2 with 
 $q_{\tau} = 0.5$ for every $\tau \in \Theta$, whereby 
 $P_{X_1 X_2 | \theta = \tau} = P_{X_1|\theta = \tau} P_{X_2 | \theta = \tau}$. 
 Clearly, $\Theta_2 = \Theta$. For a $k$-IRS, the 
 Bayesian USRDf is 
 \begin{align}
  R_{\imath}(\Delta)
  & = \underset{ \{ \Delta_{\tau}, \tau \in \Theta \} \atop \mathbbm{E}[\Delta_{\theta}] \leq \Delta } 
 \min \underset{\tau \in \Theta} \max \ \Big (  h  ( 0.5 ) - h \Big ( \frac{ \Delta_{\tau} -p_{\tau}}{1-p_{\tau}} \Big ) \Big )  \\
  & =      h   ( 0.5 ) - h \Big ( \frac{ \Delta  - p }{1- p} \Big ) 
 \end{align}
for $ 0 \leq \Delta \leq p $, where $p = \mathbbm{E}[p_{\theta}]$, and the nonBayesian USRDf is 
 \begin{align}
  R_{\imath}(\Delta)
  & =      h  ( 0.5 ) - \underset{\tau \in \Theta} \min \ 
  h \Big ( \frac{ \Delta  - p_{\tau} }{1- p_{\tau}} \Big ), 
  \quad 0 \leq \Delta \leq \underset{\tau \in \Theta} \max \ p_{\tau}.
 \end{align}
 For a $k$-MRS, in $\rho_m^{\cB}(\delta, P_U, P_{S|X_{\cM} U}, \tau)$
as well as  $\rho_m^{n\cB}(\delta, P_U, P_{S|X_{\cM} U}, \tau)$, $P_U = $ a point-mass and
 \begin{align}
\label{eq:Optimal_sampler}
 P_{S|X_{\cM} U}(s|x_{\cM},u) = P_{S|X_{\cM}}(s|x_{\cM}) 
 = \begin{cases}
      1, \ \ & s = 1, \ x_{\cM} = 00  \ \text{or} \ 11 \\
      1, \ \ & s = 2, \ x_{\cM} = 01  \ \text{or} \ 10 \\
      0, \ \ & \text{otherwise}
   \end{cases}
\end{align}
are uniformly optimal for all $ 0 \leq \delta \leq  p_{\tau} $
and for all $\tau \in \Theta$.
Then, the Bayesian USRDf is 
\begin{align}
 R_{m}(\Delta) = \underset{ \{ \Delta_{\tau}, \tau \in \Theta \} \atop \mathbbm{E}[\Delta_{\theta}] \leq \Delta } 
 \min \underset{\tau \in \Theta} \max \ \big (  h(p_{\tau}) - h( \Delta_{\tau} ) \big ),
 \quad 0 \leq \Delta \leq p
\end{align}
and the nonBayesian USRDf is 
\begin{align}
 R_{m}(\Delta) = \underset{\tau \in \Theta} \max \    h(p_{\tau}) - h( \Delta  ),
 \quad 0 \leq \Delta \leq \underset{\tau \in \Theta} \max \ p_{\tau}.
\end{align} 
 Clearly, in both the Bayesian and nonBayesian settings $R_{m}(\Delta) < R_{\imath}( \Delta)$.

 \qed 
 
\end{example}

\vspace*{0.2cm}

In closing this section, standard properties of the USRDf for the fixed-set sampler, $k$-IRS and 
$k$-MRS in the Bayesian and nonBayesian settings are summarized below, with the proof provided in
Appendix \ref{app:lemma_std_props}.

\begin{lemma} \label{l:convexity}
 The right-sides of \eqref{eq:Bayesian_kfs}, \eqref{eq:nonBayesian_kfs}, \eqref{eq:USRDf_IRS_Bayesian}, \eqref{eq:USRDf_IRS_nonBaye_fin}, \eqref{eq:USRDf-MRS-Bayesian-finite} and \eqref{eq:USRDf-MRS-nonBayesian-finite}
 are  finite-valued, decreasing, convex, continuous functions of $ \Delta_{\min} \leq \Delta \leq \Delta_{\max}$.
\end{lemma}




\section{Proofs} \label{s:Proofs}

\subsection{Achievability proofs}

Our achievability proofs emphasize the Bayesian setting. Counterpart proofs in the 
nonBayesian setting use similar sets of ideas, and so we limit ourselves to pointing
out only the distinctions between these and their Bayesian brethren. In the 
Bayesian setting, the achievability proofs successively build upon each other according
to increasing complexity of the sampler, and are presented in the order: fixed-set sampler,
$k$-IRS and $k$-MRS.

\vspace*{0.1cm}

A common theme in the achievability proofs for a $k$-FS, a $k$-IRS and a $k$-MRS involves forming estimates
 ${\widehat \tau_1}$ of the underlying $\tau_1 $ in $\Theta_1$,  ${\widehat \tau_2}$ of $\tau_2$
 in $\Theta_2$ and ${\widehat \tau}$ of $\tau$ in $\Theta$, respectively. 
 The assumed finiteness of $\Theta$ enables ${\widehat \tau_1}$ or
 ${\widehat \tau_2}$ to be conveyed rate-free to the decoder. Codes for 
 achieving USRDf at a prescribed distortion level $\Delta$ are chosen from among 
fixed-set sampling rate distortion codes for $\tau_1$s in $\Theta_1$ or from among
IRS codes for  $\tau_2$s in $\Theta_2$ or from among MRS codes for $\tau$s in $\Theta$. 
 Such codes, in the Bayesian setting, correspond to appropriate distortion
 thresholds that, in effect, average to yield a distortion level $\Delta$;
 in the nonBayesian setting, a suitable ``worst-case'' distortion must not exceed
 $\Delta$. 
 A chosen 
 code corresponds to an estimate ${\widehat \tau_1}$, ${\widehat \tau_2}$
 or ${\widehat \tau}$.
\vspace*{0.2cm}

A mainstay of our achievability proofs is the existence of sampling rate
distortion codes with fixed-set sampling for a DMMS with {\it known} pmf 
$Q$.
\begin{lemma} \label{l:ach_lemma}
 Consider a DMMS $\{X_{\cM t} \}_{t=1}^{\infty}$ with known pmf $Q = Q_{X_{\cM}}$. Let 
 $A , B \subseteq \cM$ be fixed sampling and recovery sets, respectively, and define
 \begin{align}
  d_{A}(x_{A}, y_{B})  \triangleq \mathbbm{E}[d(X_{B}, y_{B})|X_{A} = x_{A}].
 \end{align}
For every $\ep>0$  and   $\Delta_{\min} \leq \Delta \leq \Delta_{\max}$, 
there exists a sampling rate distortion code $(f, \varphi)$ of rate
\begin{align}
 \frac{1}{n} \log ||f || \leq \underset{\mathbbm{E}_{Q}[d_{A}(X_{A}, Y_{B})] 
 \leq \Delta} \min I_{Q}(X_{A} \wedge Y_{B})
 + {\ep} 
\end{align}
and expected distortion
\begin{align}
 \mathbbm{E}_{Q} \big [d \big(X_{B}^{n}, \varphi(f(X_{A}^{n}))  \big) \big]
 = \mathbbm{E}_{Q} \big [d_{A} \big(X_{A}^{n}, \varphi(f(X_{A}^{n}))  \big) \big]
 \leq \Delta + \ep
\end{align}
for all $n$ large enough.
Here, 
$$ \Delta_{\min} = \mathbbm{E}[ \underset{y_{B} \in \cY_{B}} \min d_{A}(X_{A}, y_{B})]
\ \text{ and } \ \Delta_{\max} = \underset{y_{B} \in \cY_{B}} \min \mathbbm{E}[   d_{A}(X_{A}, y_{B})].
$$
\end{lemma}
\noindent {\it Proof}: The proof of the lemma 
follows from the achievability proof of Proposition 1 
in \cite{BodNar17} upon replacing the recovery 
set $\cM$ therein by $B$.
\qed

\vspace*{0.2cm}

\noindent {\bf Theorem \ref{prop:k_FS_finite}}:
Considering first the Bayesian setting, observe that 
\begin{align}
 \Delta_{\min } 
 & = \underset{ \theta, X_{\cM} \MC \theta_{1},X_{A} \MC Y_{B} } \min \mathbbm{E}[d(X_{B}, Y_{B})] \\
 & = \underset{ \theta, X_{\cM} \MC \theta_{1},X_{A} \MC Y_{B} } \min \mathbbm{E}[ \mathbbm{E}[d(X_{B}, Y_{B})|X_{A}, \theta_{1}] ] \\
 & = \underset{ \theta, X_{\cM} \MC \theta_{1},X_{A} \MC Y_{B} } \min \mathbbm{E}[  d_{\theta_{1}}(X_{A}, Y_{B})  ] \ \ \ \  \text{by} \ \eqref{eq:modified_distortion} \\ 
 & =  \mathbbm{E}[  \mathbbm{E}[  \underset{ y_{B} \in \cY_{B} } \min d_{\theta_{1}}(X_{A}, y_{B})|\theta_{1}] ]
\end{align}
and
\begin{align}
 \Delta_{\max} & = \underset{  \theta, X_{\cM} \MC \theta_{1},X_{A} \MC Y_{B} \atop 
 P_{X_{A} Y_{B}|\theta_1 = \tau_1} =  P_{X_{A}|\theta_1 = \tau_1} P_{ Y_{B}|\theta_1 = \tau_1}, \tau_1 \in \Theta_1 } \min \mathbbm{E}[d(X_{B}, Y_{B})] \\
 & = \mathbbm{E} \big [ \underset{   
 P_{X_{A} Y_{B}|\theta_1 } = P_{X_{A}|\theta_1} P_{ Y_{B}|\theta_1}  } \min \mathbbm{E}[d_{\theta_1}(X_{A}, Y_{B})|\theta_1] \big ] \\
 & = \mathbbm{E} \big [ \underset{ y_{B} \in \cY_{B}    } \min \mathbbm{E}[d_{\theta_1}(X_{A}, y_{B})|\theta_1] \big ].
\end{align}

\vspace*{0.2cm}

Now, consider a partition $\Theta_1$ of $\Theta$ as in Section \ref{s:Results}.
Based on the sampler output  $X_{A}^{n},$ the encoder forms an ML estimate of $\theta_1$ as
$${\widehat \tau_{1,n}} = {\widehat \tau_{1,n}}(X_{A}^{n}) \triangleq \underset{ {\tau_1} \in \Theta_1 }  {\arg \max} \ P_{X_{A}^{n}|\theta_1}( X_{A}^{n}|{\tau_1}).$$ 
For each $\tau_1$ in $\Theta_1$, observe that $\{ X_{A t} \}_{t=1}^{\infty}$ 
is a DMMS with pmf $P_{\tau_1} \triangleq P_{X_{A}|\theta_1 = \tau_1}$. 
The sequence of ML estimates $\{ {\widehat \tau}_{1,n} \}_{n}$ converges in 
$P_{\tau_1}$-probability to $\tau_1$, so that for every $\ep > 0$ and $\tau_1$ in $\Theta_1$, 
there exists an $N_{1}(\ep , \tau_1)$ such that
\begin{align}
 P_{   \tau_1}( {\widehat \tau_{1,n}} \neq \tau_1 ) = P_{ \tau_1}( {\widehat \tau_{1,n}}(X_{A}^{n}) \neq \tau_1 ) \leq \frac{\ep}{2 d_{\max}}, \ \ \ n \geq N_{1}(\ep , \tau_1),
\end{align}
where $d_{\max} = \underset{x_{B} \in \cX_{B}, \ y_{B} \in \cY_{B}} \max \ d(x_{B}, y_{B})$. 
By the finiteness of $\Theta_1$, there exists an $N(\ep)$ such that 
simultaneously for all $\tau_1 \in \Theta_1$,
$$P_{ \tau_1}( {\widehat \tau_{1,n}} \neq \tau_1 ) \leq \frac{\epsilon}{2 d_{\max}}, 
\ \ n \geq N(\ep)  $$ and consequently
\begin{align} \label{eq:1k-FS-estimate-finite}
 P( {\widehat \tau_{1,n}} \neq \theta_1 ) 
 = \sum \limits_{\tau_1 \in \Theta_1} \mu_{\theta_1}(\tau_1) P_{\tau_1}( {\widehat \tau_{1,n}}  \neq \tau_1 ) 
 \leq \frac{\epsilon}{2 d_{\max}}, \ n \geq N(\ep).
\end{align}

\vspace*{0.2cm}
For a fixed $\Delta_{\min} \leq \Delta \leq \Delta_{\max}$, let $\{ \Delta_{\tau_1}, \ \tau_1 \in \Theta_1 \}$ yield  the minimum in \eqref{eq:Bayesian_kfs}. For each $\tau_1$ in $\Theta_1$, for the DMMS $\{  X_{\cM t}\}_{t=1}^{\infty}$ with pmf $P_{X_{\cM}|\theta_{1} = \tau_1} $ and distortion measure $d_{\tau_1}$, there exists by Lemma \ref{l:ach_lemma}  -- with $Q = P_{X_{\cM}|\theta_1 = \tau_1}$ and $d_{A} = d_{\tau_1}$ -- a fixed-set sampling rate distortion code  $(f_{\tau_1}, \varphi_{\tau_1}), $   
   $ f_{\tau_1}: \cX_{A}^{n} \rightarrow \{1, \ldots, J \} $ and   $\varphi_{\tau_1} : \{1, \ldots, J \} \rightarrow \cY_{B}^{n}$ of  rate $\frac{1}{n} \log J \leq  \underset{\tau_1 \in \Theta_1} \max \ \rho_{A}^{\cB}(\Delta_{{\tau_1}}, {  \tau_1}) +  \frac{\ep}{2}= R_{A}(\Delta) + \frac{\ep}{2} $ and with expected distortion
$$ \mathbbm{E}[  d_{\tau_1} ( X_{A}^{n}, \varphi_{\tau_1}( f_{\tau_1}(X_{A}^{n})) )|\theta_1 = \tau_1 ] \leq  \Delta_{\tau_1} + \frac{\ep}{2} $$ 
for all $n \geq N_{2}(\ep,\tau_1)$.

\vspace*{0.2cm}

A code $(f, \varphi)$, with $f$ taking values in $\cJ \triangleq \{1, \ldots, |\Theta_1| \} \times \{1, \ldots, J \}$ is constructed as follows. Order (in any manner) the elements of $\Theta_1$. The encoder $f$, dictated by the estimate ${\widehat \tau}_{1,n}$, is 
\begin{align}
  f(x_{A}^{n}) \triangleq ({\widehat \tau_{1,n}}(x_{A}^{n}), f_{{\widehat \tau_{1,n}}}(x_{A}^{n}) ), \ \ x_{A}^{n} \in \cX_{A}^{n}.
 \end{align}
The decoder is
\begin{align*}
  \varphi({\widehat \tau_{1,n}},j)   
  \triangleq \varphi_{ {\widehat \tau_{1,n}} } (j), \quad ({\widehat \tau_{1,n}},j) \in \cJ .
 \end{align*}

\noindent The rate of the code is 
\begin{align} \label{eq:1FS_Bayesian_finite_rate}
 \frac{1}{n} \log | \cJ|   =  \frac{1}{n} \log |\Theta_1|  + \frac{1}{n}   \log J   \leq R_{A}(\Delta)  +   \ep,
\end{align}
for all $n$ large enough, by the finiteness of $\Theta_1$.
\vspace*{0.2cm}

The code $(f,\varphi)$ is seen to satisfy
\begin{align}
 \mathbbm{E}[d(X_{B}^{n},  \varphi(f(X_{A}^{n})))]  & \leq \mathbbm{E}[ \mathbbm{1}( {\widehat \tau_{1,n}} = \theta_1 ) d(X_{B}^{n}, \varphi_{{\widehat \tau_{1,n}}}(f_{{\widehat \tau_{1,n}}}(X_{A}^{n}))) ] + P ( {\widehat \tau_{1,n}} \neq \theta_1  ) d_{\max} \\
 &   =   \mathbbm{E}[ \mathbbm{1}( {\widehat \tau_{1,n}} = \theta_1  ) d(X_{B}^{n}, \varphi_{{\theta_1}}(f_{{\theta_1}}(X_{A}^{n}))) ] + P ( {\widehat \tau_{1,n}} \neq \theta_1 ) d_{\max} \\  
 & \leq   \mathbbm{E} \big[  d(X_{B}^{n}, \varphi_{{\theta_1}}(f_{{\theta_1}}(X_{A}^{n})))   \big ] + P ( {\widehat \tau_{1,n}} \neq \theta_1 ) d_{\max} . \label{eq:1FS_Bay_eq1}
 \end{align}
 The first term on the right-side of \eqref{eq:1FS_Bay_eq1} is
 \begin{align} 
 &   \mathbbm{E} \Big[ \frac{1}{n} \sum \limits_{t=1}^{n} d  \big (X_{B {t}},  (\varphi_{{\theta_1}}(f_{\theta_1}(X_{A}^{n})))_{t} \big ) \Big]  \\
 & = \mathbbm{E} \Big[ \frac{1}{n}  \sum \limits_{t=1}^{n} \mathbbm{E} \big[  d(X_{B t}, (\varphi_{{\theta_1}}(f_{{\theta_1}}(X_{A}^{n})))_{t})|   X_{A}^{n}, \theta  \big ] \Big] \\
 & = \mathbbm{E} \Big[ \frac{1}{n}  \sum \limits_{t=1}^{n} \mathbbm{E} \big[  d(X_{B t}, (\varphi_{\theta_1}(f_{{\theta_1}}(X_{A}^{n})))_{t})|   X_{A t}, \theta  \big ] \Big],  \qquad \quad \ \text{since } P_{X_{\cM}^{n}|\theta} = \prod \limits_{t=1}^{n} P_{X_{\cM t}|\theta} \\
 & = \mathbbm{E} \Big[ \frac{1}{n}  \sum \limits_{t=1}^{n} \mathbbm{E} \big[  d(X_{B t}, (\varphi_{{\theta_1}}(f_{{\theta_1}}(X_{A}^{n})))_{t})  |   X_{A t}, \theta_1  \big ] \Big] , \qquad \quad \text{since } \theta \MC \theta_1 \MC X_{A}^{n} \\ 
 & = \mathbbm{E} \Big[ \frac{1}{n} \sum \limits_{t=1}^{n} d_{\theta_1} \big (X_{A {t}},  (\varphi_{{\theta_1}}(f_{{\theta_1}}(X_{A}^{n})))_{t} \big ) \Big],  \qquad \qquad \qquad \quad \  \text{ by }  \eqref{eq:modified_distortion}  \\
 & = \mathbbm{E}  [  d_{\theta_1}(X_{A}^{n}, \varphi_{{\theta_1}}(f_{{\theta_1}}(X_{A}^{n})))  ].  \label{eq:1FS_Bay_eq2}  
 \end{align}
 Combining \eqref{eq:1FS_Bay_eq1} and \eqref{eq:1FS_Bay_eq2},
 \begin{align} 
 \mathbbm{E}[d(X_{B}^{n},  \varphi(f(X_{A}^{n})))]  
 & \leq \mathbbm{E}  [  d_{\theta_1}(X_{A}^{n}, \varphi_{{\theta_1}}(f_{{\theta_1}}(X_{A}^{n})))  ] + P ( {\widehat \tau_{1,n}} \neq \theta_1 ) d_{\max} \\
 &   \leq \mathbbm{E} \left [\Delta_{\theta_1}  \right ] + \ep  \leq \Delta +  \ep,  \label{eq:1FS_Bayesian_finite_distortion}
\end{align}
by \eqref{eq:1k-FS-estimate-finite} for all $n$ large enough. 
Finally, we note that \eqref{eq:1FS_Bayesian_finite_rate} and \eqref{eq:1FS_Bayesian_finite_distortion} hold simultaneously for all $n$ large enough.

\vspace*{0.2cm}

In the nonBayesian setting, the achievability proof follows by adapting the steps above with the following differences. For each $\tau_1$ in $ \Theta_1$, a fixed-set sampling rate distortion code $(f_{\tau_1},\varphi_{\tau_1})$ is chosen now with expected distortion $\mathbbm{E}[  d  ( X_{B}^{n}, \varphi_{\tau_1}( f_{\tau_1}(X_{A}^{n})) )|\theta  = \tau  ]  \leq  \Delta + \frac{\ep}{2} $  for every $\tau$  in $\Lambda(\tau_1) $ and of rate   $\frac{1}{n} \log  ||f_{\tau_1}||\leq  R_{A}(\Delta) + \frac{\ep}{2} $, where  $R_{A}(\Delta)$ is the  nonBayesian USRDf for  a fixed-set sampler. 

\qed


\vspace*{0.4cm}

\noindent {\bf Theorem \ref{th:k_IRS_finite}}: 
In the Bayesian setting, for a given $\Delta_{\min} \leq \Delta \leq \Delta_{\max},$ 
consider the $P_{S}, \{ \Delta_{\tau_2} , \ \tau_2 \in \Theta_2 \}$ 
that attain the (outer) minimum in \eqref{eq:USRDf_IRS_Bayesian}. 
For the corresponding minimizing $P_{Y_{B}|S X_{S} \theta_2}$ 
in \eqref{eq:USRDf_IRS_Bayesian} (by way of \eqref{prim:IRS_bayesian})
\begin{align}
   \underset{\tau_{2} \in \Theta_{2}} \max \ \rho_{\imath}^{\cB}(\Delta_{\tau_{2}}, P_{S}, \tau_{2})  
 & = \underset{\tau_{2} \in \Theta_{2}} \max \sum \limits_{A_{i} \in \cAk} P_{S}(A_{i}) I(X_{A_{i}} \wedge Y_{B} |S = A_{i}, \theta_{2} = \tau_{2}) \label{eq:1IRS_fin_pf1}
\end{align}
and let 
\begin{align}
  \Delta_{A_{i} , \tau_{2}} \triangleq 
\mathbbm{E}[d(X_{B}, Y_{B}) | S= A_{i}, \theta_{2} = \tau_{2}] , \quad A_{i} \in \cAk, \ \tau_{2} \in \Theta_2.
\end{align}
The second expression in \eqref{eq:1IRS_fin_pf1}  
suggests an achievability scheme using an IRS code (see \cite{BodNar17}) governed by $\theta_{2}$. 
Our achievability proof comprises two phases. In the first phase an estimate ${\widehat  \tau_{2} }$ 
of $\theta_{2}$ is formed based on the output of a 
$k$-IRS that chooses each $A_i$ in $\cAk$ repeatedly for $N$ time instants. The second phase,
of length $n$, entails choosing each $S_{t} = A_{i}$ repeatedly for $ \approx nP_{S}(A_{i})$ time instants and 
an IRS code governed by ${\widehat  \tau_{2}} $ of expected distortion
\begin{align}
\sum \limits_{i} P_{S}(A_{i}) \Delta_{A_{i}, {\widehat \tau_{2}}} 
\end{align}
is applied to the output of the sampler. 
This predetermined selection of sampling sets obviates the need for the decoder to be additionally informed.

\vspace*{0.2cm}
%

Denote $|\cAk|$ by $M_{k} = {m \choose k}.$
Fix $\ep > 0$ and $0 < \ep' < \ep$. In the first phase,
a $k$-IRS is chosen 
to sample each $A_{i} \in \cAk$ over disjoint time-sets $\mu_i$
of length $N$. The union of the time-sets 
$\mu_i, \ i \in  \cM_k  \triangleq \{1, \ldots, M_{k} \}  $
is denoted by
$\mu \triangleq \{1, \ldots, M_{k}N \} $.
Based on the sampler output,
an ML estimate 
${\widehat \tau}_{2,N} = {\widehat \tau}_{2,N}(S^{\mu},X_{S}^{\mu})$ 
of $\theta_{2}$ is formed with
\begin{align} \label{eq:1estimate_IRS_fin}
 P( {\widehat \tau}_{2,N}  \neq \theta_{2} ) \leq   \frac{\ep'}{2 d_{\max}}, 
\end{align}
for $N \geq N_{\ep'}$, say.
\vspace*{0.2cm}

In the second phase, we denote the next set of $n$ time instants,
i.e., $\{M_{k}N + 1, \ldots, M_{k}N + n \}$ simply 
by $\nu \triangleq \{   1,\ldots, n  \}.$ Further, 
for each $i$ in $\cM_{k} $, define 
the time-sets $\nu_{A_{i}} \subset \nu $, made up of consecutive 
time instants, as 
\begin{align*}
 \nu_{A_{i}} = \Big \{t : 
 \lceil n \sum_{j=1}^{i-1} P_{S}(A_{j}) \rceil + 1 
 \leq t  \leq    \lceil n \sum_{j=1}^{i} P_{S}(A_{j}) \rceil  \Big   \}, 
\end{align*}
and note that the union of $\nu_{A_{i}}$s is $\nu $, and 
\begin{align}
 \left |  \frac{|\nu_{A_{i}}|}{n} - P_{S}(A_{i})  \right |
 \leq \frac{1}{n}, \ \ i \in \cM_{k}.
\end{align}

\noindent In this phase, 
the $k$-IRS is now chosen (deterministically) as follows:
\begin{align*}
 S_{t} = s_{t} = A_{i}, \ t \in \nu_{A_{i}}, \ i \in  \cM_{k}.
\end{align*}

\vspace*{0.2cm}

For each DMMS $\{ X_{\cM t} \}_{t=1}^{\infty}$ 
with pmf $P_{X_{\cM}|\theta_{2} = \tau_{2}}, \ \tau_2 \in \Theta_2$,
and for each $A_{i}$ in $\cAk$ and its corresponding 
distortion measure $d_{\tau_{2}},$ there exists by Lemma \ref{l:ach_lemma}
-- with $Q = P_{X_{\cM}|\theta_{2} = \tau_{2}}$ and
$d_{A} = d_{\tau_2}$ --
 a fixed-set sampling rate distortion code 
$(f_{A_{i}}^{ \tau_{2}}, \varphi_{A_{i}}^{ \tau_{2}}), \ 
f_{A_{i}}^{ \tau_{2}}: \cX_{A_{i}}^{ \nu_{A_{i}} } 
\rightarrow \{1, \ldots , J_{A_{i}}^{\tau_{2}} \} $ and 
$\varphi_{A_{i}}^{\tau_{2}} : \{1, \ldots , J_{A_{i}}^{\tau_{2}} \} 
\rightarrow \cY_{B}^{\nu_{A_{i}}} $ of rate 
$\frac{1}{| \nu_{A_{i}} |} \log J_{A_{i}}^{\tau_{2}} 
\leq I(X_{A_{i}} \wedge Y_{B}|S = A_{i}, \theta_{2} = {\tau_{2}})
+ \frac{\ep'}{4}$ (cf. \eqref{eq:1IRS_fin_pf1}) and with
\begin{align} 
 \mathbbm{E} \Big[ d_{\tau_{2}} \big ( X_{A_{i}}^{\nu_{A_{i}}}, 
 \varphi_{A_{i}}^{\tau_{2}}(f_{A_{i}}^{\tau_{2}}(X_{A_{i}}^{\nu_{A_{i}}} ) ) \big )
 \big | \theta_{2}= \tau_{2} \Big ]  \leq \Delta_{A_{i}, \tau_{2}} +\frac{\ep'}{2},
\end{align}
for all $ |\nu_{A_{i}}|  \geq N_{A_{i}}   ( {\ep'}, \tau_{2} ).$ 
Note that 
$$\sum \limits_{\tau_{2} \in \Theta_{2}} \mu_{\theta_{2}}(\tau_{2}) 
\sum \limits_{ i = 1}^{M_{k}}  P_{S}(A_{i})  \Delta_{A_{i}, \tau_{2}}   \leq \Delta$$   
and $$\sum \limits_{i=1}^{M_{k}} P_{S}(A_{i}) 
I(X_{A_{i}} \wedge Y_{B}|S = A_{i}, \theta_{2}= \tau_{2}) 
\leq R_{\imath}(\Delta)$$ for every $\tau_{2}$ in $\Theta_{2}$.

\vspace*{0.2cm}

Consider a (composite) code $(f, \varphi)$ as follows. 
Denote $n' \triangleq  |\mu| + |\nu| = M_{k} N + n,$
and the encoder $f$ consisting of a concatenation of 
encoders is defined by
\begin{align}
 f(s^{n'}, x^{n'}) \triangleq \Big ({\widehat \tau}_{2,N},
 f_{A_{1}}^{{\widehat \tau}_{2,N}}(x_{A_{1}}^{\nu_{A_1}}),
 \ldots, f_{A_{M_{k}}}^{{\widehat \tau}_{2,N}}(x_{A_{M_{k}}}^{\nu_{A_{M_{k}}}}) \Big ).
\end{align}
The decoder $\varphi$, which is aware of the predetermined sequence of sampling sets,
is defined by
\begin{align}
\varphi  (s^{n'},  {\widehat \tau}_{2,N}, j_{1}, \ldots, j_{M_{k}}) =  \varphi ( {\widehat \tau}_{2,N}, j_{1}, \ldots, j_{M_{k}})  \triangleq 
 \big ( \underbrace{ y_{1}, \ldots, y_{1}}_{ \text{first phase}},
 \underbrace{\varphi_{A_{1}}^{ {\widehat \tau}_{2,N}} (j_{1}),
 \ldots, \varphi_{A_{M_{k}}}^{ {\widehat \tau}_{2,N}} 
 (j_{M_{k}})}_{\text{second phase}}   \big ),
\end{align}
for each encoder output $({\widehat {{\tau}}}_{2,N}, j_{1}, \ldots, j_{M_{k}})$, 
where  $y_{1} \in \cYm$ is an arbitrary symbol.  Clearly, 
  $ |\Theta_{2}| \times \underset{\tau_{2} \in \Theta_{2}}
\max   \prod  \limits_{i=1}^{M_{k}}  J_{A_{i}}^{{  \tau_{2}} }$ 
indices would suffice to describe all possible encoder outputs.

\vspace*{0.2cm}

The rate of the code is  
\begin{align}
 \!    \frac{1}{n'} \log |\Theta_{2}| \! + 
 \! \underset{\tau_{2} \in \Theta_{2}} \max \ 
 \frac{1}{n'} \!  \sum \limits_{i=1}^{M_{k}} \! 
 \log J_{A_{i}}^{{  \tau}_{2}} \!   & \leq
 \underset{\tau_{2} \in \Theta_{2}} \max \sum \limits_{i=1}^{M_{k}} 
 \frac{|\nu_{A_{i}}|}{n} \frac{1}{|\nu_{A_{i}}|} 
 \log J_{A_{i}}^{{  \tau}_{2}} + \frac{1}{n'} \log |\Theta_{2}| \!  \\
& \! \leq  \underset{\tau_{2} \in \Theta_{2}} \max 
\sum \limits_{i=1}^{M_{k}}\!  \Big ( \! P_{S}(A_{i}) \!
+ \! \frac{1}{n} \Big ) \! \Big (\!I(X_{A_{i}} \wedge Y_{B}| 
S = A_{i}, \! \theta_{2} = \tau_{2}) \! + \! \frac{\ep'}{4} \Big ) \! 
+ \! \frac{1}{n'} \log |\Theta_{2}|  \\
& \leq \underset{\tau_{2}  \in \Theta_{2}} \max 
\sum \limits_{i=1}^{M_{k}} P_{S}(A_{i}) I(X_{A_{i}} \wedge Y_{B}|S = A_{i}, 
\theta_{2} = \tau_{2}) +  \ep'  < R_{\imath}(\Delta) +  \ep , \label{eq:1IRS_Bayesian_finite_rate}
\end{align}
where the previous inequality holds for all $n$ large enough. 
Denoting the 
output of the decoder by $Y_{B}^{n'} \triangleq 
\varphi(f(S^{n'}, X_{S}^{n'}))$  
\begin{align}
\! \mathbbm{E}[d(X_{B}^{n'}, Y_{B}^{n'})] \!
=   \frac{1}{n' } \mathbbm{E} \Big [   \sum \limits_{t \in \mu } 
d(X_{B t}, Y_{B t}) \! + \! \sum \limits_{t \in \nu } \! 
\big ( \mathbbm{1}( {\widehat \tau}_{2,N} \neq \theta_{2} )d(X_{B t}, Y_{B t}) +  
\mathbbm{1}( {\widehat \tau}_{2,N} = \theta_{2} ) 
d(X_{B t}, Y_{B t}) \big ) \Big ]. \ \ \ \ \label{eq:1IRS_fin_dist_p1}
\end{align}
\noindent The first two terms on the right-side of \eqref{eq:1IRS_fin_dist_p1} are
\begin{align}
 \mathbbm{E} \Big [  \frac{1}{n'} \sum \limits_{t \in  \mu }  d(X_{B t}, Y_{B t}) 
 + \frac{\mathbbm{1}( {\widehat \tau}_{2,N} \neq \theta_{2} )}{n' } 
 \sum \limits_{t \in \nu } d(X_{B t}, Y_{B t}) \Big ] \leq 
 \frac{ M_{k} N d_{\max} }{n'} + \frac{\ep'}{2}, \label{eq:1IRS_fin_dist_p2}
\end{align} 
by \eqref{eq:1estimate_IRS_fin} for $N$ large enough, and the last term
on the right-side of \eqref{eq:1IRS_fin_dist_p1} is
\begin{align}
 \mathbbm{E} \Big [ \frac{\mathbbm{1}( {\widehat \tau}_{2,N} = \theta_{2} )}{n' }
 \! \sum \limits_{t \in \nu } d(X_{B t}, Y_{B t})  \Big ]
 & \leq  \sum \limits_{i=1}^{M_{k}} \frac{|\nu_{A_{i}}|}{n} 
 \mathbbm{E} \Big [ \mathbbm{1}({\widehat \tau}_{2,N} = \theta_{2} ) 
 d \big (X_{B}^{\nu_{A_{i}}}, \varphi_{{A_{i}}}^{{\widehat \tau}_{2,N}} 
 ( f_{{{A_{i}}}}^{{\widehat \tau}_{2,N}} (X_{A_{i}}^{\nu_{A_{i}}} ) ) \big) \Big ] \\
 & \leq \sum \limits_{i=1}^{M_{k}} \frac{|\nu_{A_{i}}|}{n} 
 \mathbbm{E} \big [ d \big (X_{B}^{\nu_{A_{i}}}, \varphi_{{A_{i}}}^{\theta_{2}} 
 ( f_{{{A_{i}}}}^{\theta_{2}} (X_{A_{i}}^{\nu_{A_{i}}} ) ) \big)   \big ]   \\
 & = \sum \limits_{i=1}^{M_{k}} \frac{|\nu_{A_{i}}|}{n}
 \mathbbm{E} \big [ d_{\theta_{2}} \big (X_{A}^{\nu_{A_{i}}}, 
 \varphi_{{A_{i}}}^{\theta_{2}}  ( f_{{{A_{i}}}}^{\theta_{2}} (X_{A_{i}}^{\nu_{A_{i}}} ) ) \big)   \big ] \\
 & \leq \sum \limits_{i=1}^{M_{k}} \Big  (P_{S}(A_{i}) 
 + \frac{1}{n} \Big) \mathbbm{E} \big [  \Delta_{A_{i}, \theta_{2}} 
 + \frac{\ep'}{2}  \big ] \\ 
 & \leq \Delta + \frac{\ep'}{2} +   \frac{1}{n} \sum \limits_{i=1}^{M_{k}} \mathbbm{E}
 [  \Delta_{A_{i}, \theta_{2}}  ]  
 + \frac{M_{k}}{n} \frac{\ep'}{2}. \label{eq:1IRS_fin_dist_p3}  
\end{align}
From  \eqref{eq:1IRS_fin_dist_p1}-\eqref{eq:1IRS_fin_dist_p3}, we have
\begin{align}
\mathbbm{E}[d(X_{B}^{n'}, Y_{B}^{n'})] \leq \Delta + \ep, \label{eq:1IRS_Bayesian_finite_distortion}
\end{align}
for $n$ and $N$ large enough. Finally, we note that \eqref{eq:1IRS_Bayesian_finite_rate} and \eqref{eq:1IRS_Bayesian_finite_distortion} hold simultaneously for all $n$ and $N$ large enough.

\vspace*{0.1cm}

\noindent The Corollary is immediate by the choice of codes with ``uninformed'' decoder in the 
proof above.

\vspace*{0.2cm}

For the nonBayesian setting, achievability follows by adapting the proof above 
in a manner similar to that for a $k$-FS in Theorem \ref{prop:k_FS_finite}.

\qed

\noeqref{eq:1IRS_fin_dist_p2}


\vspace*{0.1cm}

\noindent {\bf Theorem \ref{th:k_MRS_finite}}: 
The achievability proof relies on the deterministic sampler justified by 
Proposition \ref{prop:MRS_fin_opt_samp}.
In the Bayesian setting, for a given  
$\Delta_{\min} \leq \Delta \leq \Delta_{\max}$, let
$P_{U}, \ P_{S|X_{\cM} U} = \delta_{w  }, \  \{ \Delta_{\tau}, \ \tau \in \Theta  \} $ 
 attain the minimum in \eqref{eq:USRDf-MRS-Bayesian-finite_alt}. 
For the corresponding minimizing $P_{Y_{B}|S X_{S} U \theta}$ in \eqref{prim:MRS_Bayesian}, 
 the right-side of \eqref{eq:USRDf-MRS-Bayesian-finite_alt} is
\begin{align}
\underset{\tau \in \Theta} \max \ \rho_{m}^{\cB}(\Delta_{\tau}, P_{U},  \delta_{w  }, \tau)  
 & =  \underset{\tau \in \Theta} \max \sum \limits_{u \in \cU}  P_{U}(u) I(X_{S} \wedge Y_{B} | S, U=u, \theta = \tau ) \label{eq:MRS_Bayes_fin_eq1}
\end{align}
and we set 
$$\Delta_{A_{i},u,\tau} \triangleq \mathbbm{E}[d(X_{B},Y_{B})| S = A_{i}, U = u, \theta = \tau], \quad A_{i} \in \cAk, \ \tau \in \Theta , \   u \in \cU.  $$ 
Our achievability proof uses a $k$-MRS in two distinct modes. 
First, a deterministic $k$-MRS is chosen so as to form an 
estimate ${\widehat \tau}$ of $\theta$ from the sampler output.
Next, for each $U=u$, a suitable deterministic $k$-MRS is chosen in accordance with
$w(x_{\cM},u)$,  and 
an MRS code (see \cite{BodNar17}) governed by ${\widehat \tau}$ of expected distortion
\begin{align}
  \overset{\vspace*{0.2cm} \sim \vspace*{-0.24cm} } \leq  \   
    \sum \limits_{ A_{i} } P_{S|U \theta}(A_{i}|u, {\widehat \tau}) \Delta_{A_{i},u,{\widehat \tau}}  
\end{align}
is applied to the sampler output.
Concatenation of such codes corresponding to various $u \in \cU$ yields, in effect,
time-sharing that serves to achieve \eqref{eq:MRS_Bayes_fin_eq1}. 
To simplify the notation, the conditioning on $U=u$ will 
be suppressed except when needed.

\vspace*{0.2cm}
Fix $\ep > 0$ and $0 < \ep' < \ep$. 

(i) We devise a
 deterministic $k$-MRS on a time-set $\mu  $, 
based on whose output an estimate 
${\widehat \tau}_{N} = {\widehat \tau}_{N}(S^{\mu}, X_{S}^{\mu}) = {\widehat \tau}_{N}(S^{\mu} )$ of $\theta$ is formed with
\begin{align} \label{eq:MRS_fin_estimate}
 P( {\widehat \tau}_{N} \neq \theta) \leq \frac{\ep'}{4 d_{\max}},
\end{align}
for $N \geq N_{\ep'}$. The estimate ${\widehat \tau}_{N}$ is formed
from {\it only the sampling sequence} $S^{\mu}$ and thus is available to
the encoder as well as the decoder.  The $k$-MRS is chosen on the
time-set
 $\mu$,
to signal the occurrences of each $x  \in \cXm$ to the encoder and decoder 
through $S^{\mu}$ above; for each $x  \in \cXm,$ a distinct
$A \in \cAk$ is chosen. If $|\cAk| \geq |\cXm|$, a trivial 
one-to-one mapping from $\cXm$ to $\cAk$ enables $S^{\mu}$ to determine
$X_{\cM}^{\mu}$, where $S^{\mu}$ is of length $N$, say. Then ${\widehat \tau}_{N}$ is taken to be the ML
estimate of $\theta$ based on $X_{\cM}^{\mu}$, which satisfies \eqref{eq:MRS_fin_estimate}.

\vspace*{0.1cm}

When $|\cAk| < |\cXm|,$ a $k$-MRS is chosen attuned variously to disjoint subsets of 
$\cXm$, of size $|\cAk|-1$, on corresponding disjoint time-sets $\mu_l$ of length 
$N$, $l = 1, \ldots,  \left \lceil \frac{|\cXm|}{ |\cAk| - 1} \right   \rceil$, as follows. In each
$\mu_l$, the $k$-MRS signals the occurrence (or not) of $X_{\cM t} = x $ in the $l^{th}$-subset
of $\cXm$ in a (deterministic) manner by choosing $|\cAk|-1$ distinct sampling sets
in $\cAk$; the nonoccurrence  of symbols from this $l^{th}$-subset of $\cXm$ is
indicated by the remaining (dummy) sampling set in $\cAk$. We denote 
$ \bigcup \limits_{l} \mu_l $ by $\mu$. Finally, ${\widehat \tau}_{N}$ is taken as the 
ML estimate of $\theta$ based on the sampling sequence $S^{\mu}$ of length 
$ \left \lceil \frac{|\cXm|}{ |\cAk| - 1}  \right \rceil N =   N'$, say.

\vspace*{0.2cm}

(ii) Next, for each $U=u$, a $k$-MRS is chosen according to
 $P_{S|X_{\cM}, U=u} =  \delta_{w (\cdot,u)}$  
for $n$ time instants. 
Then, for a DMMS $\{X_{\cM t} \}_{t=1}^{\infty}$ with pmf $P_{X_{\cM}|\theta = {\widehat \tau}_{N}}$ an MRS code comprising a concatenation of fixed-set sampling
rate distortion codes corresponding to the $A_{i}$s in $\cAk$ is applied
to the sampler output.

\vspace*{0.2cm}

Denote the set of $n$ time instants  
$\{  N' + 1, \ldots,  N' + n \}$ simply by 
$\gamma \triangleq \{ 1, \ldots, n\}.$  Define time-sets $ {\gamma_{S^{n}} (A_{i})} \triangleq \{t: 1 \leq t \leq  n, \ S_{t} = A_{i} \}, \ i \in \cM_{k},$
and note that $ {\gamma_{S^{n}} (A_{i})} $s cover $\gamma $,  i.e.,
\begin{align}
 \gamma = \bigcup \limits_{A_{i} \in \cAk} \gamma_{S^{n}}(A_{i}).
\end{align}
Denote the set of the first $ \max \{ \lceil (n P_{S|\theta}(A_{i}|{\widehat \tau}_{N})) - \ep' \rceil ,0 \}$ time instants in each $\gamma_{S^{n}}(A_{i})$ by $\nu_{A_{i}}$ (suppressing the dependence on ${\widehat \tau}_{N}$). 
Defining the (typical) set for each $\tau$ in $\Theta$
\begin{align}
 {\cal T}^{(n)}(\ep', {\tau} ) \triangleq  \left \{ s^{n} \in \cAk^{n}: \Big | \frac{ |\gamma_{s^{n}}(A_{i})|}{n} - P_{S|\theta}(A_{i}|{\tau}) \Big | \leq \ep', \ i \in \cM_{k} \right \},
\end{align}
we have that 
\begin{align} \label{eq:MRS_Bayes_fin_ac_typ}
P(S^{\gamma}  \notin {\cal T}^{(n)}(\ep', {\widehat \tau}_{N})) = P( S^{\gamma} \notin {\cal T}^{(n)}(\ep',{\widehat \tau}_{N}), \  {\widehat \tau}_{N} = \theta ) + P( S^{\gamma} \notin {\cal T}^{(n)}(\ep',{\widehat \tau}_{N}), \  {\widehat \tau}_{N} \neq \theta ) \leq \frac{\ep'}{2 d_{\max}} 
\end{align}
for all $n$ large enough.

\vspace*{0.2cm}

By Lemma \ref{l:ach_lemma}, for {\it each} DMMS $\{X_{\cM t} \}_{t=1}^{\infty}$ 
with pmf $P_{X_{\cM}|S = A_{i},\theta = \tau}, \ i \in \cM_{k}, \ \tau \in \Theta,$ 
there exists a code $(f_{A_{i}}^{\tau}, \varphi_{A_{i}}^{\tau}), \ 
f_{A_{i}}^{\tau}: \cX_{A_{i}}^{\nu_{A_{i}}} \rightarrow \{1, \ldots, 
J_{A_{i}}^{\tau} \} $ and $\varphi_{A_{i}}^{\tau}: \{1, \ldots, 
J_{A_{i}}^{\tau} \} \rightarrow \cY_{B}^{\nu_{A_{i}}}$ of rate 
\begin{align} \label{eq:MRS_Bayes_fin_ac_rateA}
\frac{1}{|\nu_{A_{i}}|} \log J_{A_{i}}^{\tau} \leq I(X_{A_{i}} 
\wedge Y_{B} | S = A_{i}, \theta = \tau) + \frac{\ep'}{2} 
\end{align}
and with
\begin{align} \label{eq:MRS_Bayes_fin_ac_distA}
 \mathbbm{E} \left [ d \big (X_{B}^{\nu_{A_{i}}},
 \varphi_{A_{i}}^{\tau}( f_{A_{i}}^{\tau}(X_{A_i}^{\nu_{A_{i}}}) ) \big )\big | 
 S^{\nu_{A_{i}}} = A_{i}^{\nu_{A_{i}}}, \theta = \tau \right ] 
 \leq \Delta_{A_{i},\tau} + \frac{\ep'}{4}
\end{align}
for all $|\nu_{A_{i}}| \geq N_{A_{i}}( {\ep'} , \tau )$. Such codes are 
considered for each $U=u.$

\vspace*{0.2cm}
Consider a (composite) code $(f,\varphi)$ as follows. 
Denoting $N' + n$ by $n'$, an encoder $f$ consisting of a concatenation of encoders
 is defined as 
\begin{align}
 f(s^{n'}  , x_{s}^{n'})   \triangleq  \begin{cases}
                                               \big (   f_{A_{1}}^{{\widehat \tau}_{N}}(x_{A_{1}}^{\nu_{A_{1}}}), \ldots, f_{A_{M_{k}}}^{{\widehat \tau}_{N}}(x_{A_{M_{k}}}^{\nu_{A_{M_{k}}}}) \big)  , & \   s^{\gamma}  \in {\cal T}^{(n)}(\ep', {\widehat \tau}_{N} )  \\
                                            (    1, \ldots,1), &  \ s^{\gamma}  \notin {\cal T}^{(n)}(\ep', {{\widehat \tau}_{N}} ).
                                           \end{cases}                                           
\end{align}
For $t = 1 ,\ldots, n',$ and each encoder output $( j_{1}, \ldots, j_{M_{k}}) $, the decoder $\varphi$, which 
can recover the estimate ${\widehat \tau}_{N}$ from its knowledge of the sampling sequence $S^{n'} = s^{n'}$,
is given by
\begin{align}
 \Big ( \varphi (s^{n'},  j_{1}, \ldots, j_{M_{k}} ) \Big)_{t} \triangleq
 \begin{cases}
  \Big ( \varphi_{A_{i}}^{ {\widehat \tau}_{N}} (j_{i})    \Big )_{t}, \ & s^{ \gamma} \in {\cal T}^{(n)}({\ep'}, {\widehat \tau}_{N} ) \text{ and } t \in \nu_{A_{i}}, \ i \in \cM_{k} \\
  y_{1}, & \text{ otherwise},
 \end{cases}
\end{align}
where $y_{1}$ is a fixed but arbitrary symbol in $\cYm.$
 
\vspace*{0.2cm}

Finally, for $N$ and $ n$ large enough, the codes $(f,\varphi)$  
corresponding to each $U = u$ are concatenated so as to effect the 
time-sharing prescribed by $P_{U}$, in a standard manner. It is
shown in Appendix \ref{app:MRS_rate_dis} that the rate of the resulting code is
\begin{align}
  & \substack{\sim \\ \leq \\\ }  \ \underset{\tau \in \Theta} \max \sum \limits_{u \in \cU} P_{U}(u)  
  \sum \limits_{A_{i} \in \cAk}  P_{S|U \theta}(A_{i}|u,\tau) 
   I( X_{A_{i}} \wedge Y_{B} | S = A_{i}, U = u , \theta = \tau ) + \ep' \\ 
 & \substack{\sim \\ \leq \\\ } \ R_{m}(\Delta) + \ep, \label{eq:MRS_rate_ac}
\end{align}
using \eqref{eq:MRS_Bayes_fin_ac_rateA} 
and the expected distortion is
\begin{align}
 & \substack{\sim \\ \leq \\\ } \ \mathbbm{E}[ \Delta_{S, U,  \theta} ] + \ep \\
 &\substack{\sim \\ \leq \\\ } \ \Delta + \ep, \label{eq:MRS_dist_ac}
\end{align}
 from \eqref{eq:MRS_fin_estimate}, \eqref{eq:MRS_Bayes_fin_ac_typ}, \eqref{eq:MRS_Bayes_fin_ac_distA}
 and the definition of $\Delta_{A_{i},u,\tau}$.
\qed

\vspace*{0.3cm}

\subsection{Converse proof}

In contrast with the achievability proofs, we present a 
unified converse proof for Theorems \ref{th:k_MRS_finite},
\ref{th:k_IRS_finite} and \ref{prop:k_FS_finite} according to 
successive weakening of the sampler, viz. $k$-MRS, $k$-IRS and 
fixed-set sampler. We begin with the technical Lemma \ref{l:markov_single_letter}
that is used subsequently in the converse proof.

\begin{lemma} \label{l:markov_single_letter}
Let finite-valued rvs $C, D^{n}, E^{n}, F^{n},$ be such that $(D_{t}, E_{t}), \ t = 1, \ldots,n,$ are conditionally mutually independent given $C$, i.e.,   
\begin{align} \label{eq:markov_lemma_eq1}
\hspace*{2cm}  P_{D^{n} E^{n}|C} = \prod \limits_{t=1}^{n} P_{D_{t} E_{t}|C}
\end{align}
and   satisfy
\begin{align} \label{eq:markov_lemma_eq2}
 C, D^{n} \MC E^{n} \MC F^{n}.
\end{align}
For any function $g(C)$ of $C$, such that  
\begin{align} \label{eq:markov_lemma_eq3}
E^{n} \MC g(C) \MC C \quad  \text{ and } \quad P_{E^{n}|g(C)} = \prod \limits_{t=1}^{n} P_{E_{t}|g(C)},
\end{align}
it holds that 
 \begin{align} \label{eq:markov_lemma_eq4}
\qquad  C, D_{t} \MC g(C), E_{t} \MC F_{t}, \quad t = 1, \ldots,n.
 \end{align}
\end{lemma}

\noindent {\bf Proof}: First, from  \eqref{eq:markov_lemma_eq2}, we have
\begin{align}
 0 = I(C, D^{n} \wedge F^{n} |E^{n}) & = I \big (C \wedge F^{n} |E^{n} \big ) + I \big (D^{n} \wedge F^{n} |E^{n}, C \big ) \\
 & = I \big (C, g(C) \wedge F^{n} |E^{n} \big ) + I \big (D^{n} \wedge F^{n} |E^{n}, C \big ) \\
 & \geq  I(C \wedge F^{n} | E^{n}, g(C)) + I(D^{n} \wedge F^{n} |E^{n},C). \label{eq:markov_lemma_eq5}
\end{align}
Now, the second term on the right-side of  \eqref{eq:markov_lemma_eq5} is 
\begin{align}
 0 =  I(D^{n} \wedge F^{n} |E^{n},C) & = H(D^{n} |E^{n}, C) - H(D^{n}|E^{n}, F^{n},C)\\
 & = \sum \limits_{t=1}^{n} \left ( H(D_{t} | E_{t},C) - H(D_{t} | D^{t-1}, E^{n}, F^{n},C) \right ), \ \ \ \ \text{ by } \eqref{eq:markov_lemma_eq1} \\
 & \geq \sum \limits_{t=1}^{n} \big ( H(D_{t}|E_{t},C) - H(D_{t}|E_{t}, F_{t}, C) \big ) \\
 & = \sum \limits_{t=1}^{n} I(D_{t} \wedge F_{t} |E_{t}, C). \label{eq:markov_lemma_eq6}
\end{align}
Next, the first part of
\eqref{eq:markov_lemma_eq3} along with   \eqref{eq:markov_lemma_eq5} 
implies that
\begin{align}
0  & = I \big (C \wedge E^{n} | g(C) \big ) + I \big (C \wedge F^{n} |E^{n}, g(C) \big )  \\
 & = I \big (C \wedge E^{n}, F^{n} | g(C) \big ),  
\end{align}
and hence 
\begin{align}
 I \big (C \wedge E_{t}, F_{t} | g(C) \big ) = 0, \ \ t = 1, \ldots,n. \label{eq:markov_lemma_eq7}
\end{align}
Now, by \eqref{eq:markov_lemma_eq6} and \eqref{eq:markov_lemma_eq7}, for $t = 1, \ldots, n,$
\begin{align}
I \big (C, D_{t} \wedge F_{t} | E_{t}, g(C) \big ) = I \big( C \wedge F_{t} |E_{t}, g(C) \big)  + I \big(D_{t} \wedge F_{t}| E_{t}, C\big) = 0,
\end{align}
which is the claim \eqref{eq:markov_lemma_eq4}.

\qed



\vspace*{0.3cm}

\noindent {\bf Converse}: In the Bayesian setting, we provide first a converse proof for Theorem \ref{th:k_MRS_finite}, which is then refashioned to give converse proofs for Theorems \ref{th:k_IRS_finite} and \ref{prop:k_FS_finite}. 

\vspace*{0.2cm}

Let $(\{P_{S_{t}|X_{\cM t} \theta} = P_{S_{t} | X_{\cM t} } \}_{t=1}^{\infty},
f, \varphi)$ be an  $n$-length $k$-MRS block code of rate $R$ and with 
decoder output $Y_{B}^{n} = \varphi(S^{n}, f(S^{n}, X_{S}^{n}))$  satisfying 
$\mathbbm{E}[d(X_{B}^{n}, Y_{B}^{n})] \leq \Delta.$ The hypothesis of Lemma \ref{l:markov_single_letter}
is met with $C = \theta, \ D^{n} = X_{\cM}^{n}, \ E^{n} = (S^{n}, X_{S}^{n}),
\ F^{n} =  Y_{B}^{n}$ and $g(\theta) = \theta$, since
\begin{align}
 & P_{X_{\cM}^{n} S^{n} |\theta} = P_{X_{\cM}^{n}|\theta} P_{ S^{n} |X_{\cM}^{n}} =  \prod \limits_{t=1}^{n} P_{ X_{\cM t}|\theta}  P_{ S_t | X_{\cM t}} =  \prod \limits_{t=1}^{n} P_{ X_{\cM t} S_t|\theta}  , 
  \label{eq:conv_finite_hyp3}
\end{align}
while
\begin{align}
\theta, X_{\cM}^{n} \MC S^{n}, X_{S}^{n} \MC Y_{B}^{n} \label{eq:conv_finite_hyp1}
\end{align}
holds by code construction. Also, \eqref{eq:conv_finite_hyp3} implies, upon summing
over all realizations of $X_{S^{c}}^{n}$, that
\begin{align} \label{eq:conv_finite_hyp5}
 P_{S^{n} X_{S}^{n}|\theta} = \prod \limits_{t=1}^{n} P_{S_t  X_{S_{t}} | \theta}.
\end{align}
Then the claim of the lemma implies that
\begin{align} \label{eq:conv_finite_Markov}
 \theta, X_{\cM t} \MC \theta , S_{t}, X_{S_{t}} \MC Y_{B t}, \ \ \ t=1, \ldots,n.
\end{align}
Let $ \Delta_{\tau }$ denote $\mathbbm{E}[d(X_{B}^{n}, Y_{B}^{n})|\theta  = \tau ] = \frac{1}{n} \sum \limits_{t=1}^{n} \mathbbm{E}[d(X_{B t} , Y_{B t} )|\theta  = \tau ] $ for each $\tau$ in $\Theta$ 
and note that $\mathbbm{E}[\Delta_{\theta}] \leq \Delta$. For every $\tau$ in $\Theta $,  the following holds:
\begin{align}
 R = \frac{1}{n} \log || f|| & \geq \frac{1}{n} H( f( S^{n}, X_{S}^{n})|\theta = \tau ) \geq \frac{1}{n} H( f( S^{n}, X_{S}^{n})| S^{n}, \theta  = \tau ) \\
 & \geq \frac{1}{n} H( \varphi(S^{n},f(S^{n},X_{S}^{n}))| S^{n}, \theta  = \tau ) = \frac{1}{n} H(Y_{B}^{n} | S^{n}, \theta = \tau )\\
 & = \frac{1}{n} I(X_{S}^{n} \wedge Y_{B}^{n} | S^{n}, \theta  = \tau ) \\
 & = \frac{1}{n} \sum \limits_{t=1}^{n} \left (  H( X_{S_{t}}| S^{n}, X_{S}^{t-1}, \theta  = \tau ) - H( X_{S_{t}}| S^{n}, X_{S}^{t-1}, Y_{B}^{n}, \theta  = \tau  ) \right ) \\
 & \geq \frac{1}{n} \sum \limits_{t=1}^{n} \left (  H( X_{S_{t}}| S^{n}, X_{S}^{t-1}, \theta  = \tau ) - H( X_{S_{t}}| S_{t}, Y_{B t}, \theta  = \tau  ) \right ) \\
 & =  \frac{1}{n} \sum \limits_{t=1}^{n} \left (  H( X_{S_{t}}| S_{t}, \theta  = \tau ) - H( X_{S_{t}}| S_{t}, Y_{B t}, \theta  = \tau  )  \right ),  \qquad \qquad \text{ by } \eqref{eq:conv_finite_hyp5} \\
 & =  \frac{1}{n} \sum \limits_{t=1}^{n}  I(X_{S_t} \wedge  Y_{B t} | S_{t}, \theta  = \tau ). \label{eq:conv_finite_eq1} 
\end{align}
\noindent By \eqref{eq:conv_finite_Markov},  
\begin{align} \label{eq:conv_combo}
\Big ( \big(\frac{1}{n} \sum \limits_{t=1}^{n} \mathbbm{E}[d(X_{B t} , Y_{B t} )|\theta  = \tau ] , \frac{1}{n} \sum \limits_{t=1}^{n}  I(X_{S_t} \wedge  Y_{B t} | S_{t}, \theta  = \tau )  \big),  \tau \in \Theta  \Big )
\end{align}
lies in the convex hull of  
\begin{align}
\! {\cal C} \! \triangleq \! \big \{  \big ( (\mathbbm{E}[d(X_{B}, Y_{B})|\theta = \tau ],    I(X_{S} \wedge Y_{B}|S,  \theta = \tau)),   \tau \in \Theta \big) :    P_{\theta X_{\cM} S Y_{B}} \! = \! \mu_{\theta}  P_{X_{\cM}|\theta} P_{S|X_{\cM} } P_{Y_{B}|S X_{S}  \theta} \big \} \! \subset \! \mathbbm{R}^{2|\Theta| }.   \!
\end{align}
By the Carath\'eodory Theorem \cite{CoverThomas}, every point in 
the convex hull of $ {\cal C}  $ can be represented as a convex 
combination of at most $2|\Theta| + 1$ elements in ${\cal C}$. The 
corresponding pmfs are indexed by the values of a rv $U$ with
\begin{align} \label{eq:conv_joint_pmf}
P_{U \theta  X_{\cM} S Y_{B}} =  P_{U} \mu_{\theta} P_{X_{\cM}|\theta} 
P_{S|X_{\cM} U} P_{Y_{B}|S X_{S}  \theta U}  ,
\end{align}
where the pmf of $U$ has support of size $\leq 2 |\Theta| + 1$.
Then, in a standard manner, \eqref{eq:conv_finite_eq1} leads to
\begin{align}
 R & \geq \underset{   P_{Y_{B}|S X_{S} U, \theta = \tau} \atop 
 \mathbbm{E}[d(X_{B}, Y_{B})| \theta = \tau] \leq \Delta_{\tau} } \min I(X_{S} \wedge Y_{B} |S, U, \theta = \tau) \label{eq:conv_finite_eq5}\\ 
 &  = \rho_{m}^{\cB}(\Delta_{\tau},P_{U}, P_{S|X_{\cM}U}, \tau). \label{eq:conv_finite_eq4}  
\end{align}
Now, \eqref{eq:conv_finite_eq4} holds for every $\tau \in \Theta$, and hence
\begin{align}
 R & \geq \underset{\tau \in \Theta} \max \ \rho_{m}^{\cB}(\Delta_{\tau},P_{U}, P_{S|X_{\cM}U}, \tau)
 \label{conv:finite_eq2}  \\
  & \geq \underset{P_{U}, P_{S|X_{\cM} U}, \{ \Delta_{\tau}, \ \tau \in \Theta \} \atop \mathbbm{E}[\Delta_{\theta}] \leq \Delta } \min \ \underset{\tau \in \Theta}   \max \ \rho_{m}^{\cB}(\Delta_{\tau},P_{U}, P_{S|X_{\cM}U}, \tau)
\label{conv:finite_eq3}  \\
  & = R_{m}(\Delta)
\end{align}
for $  \Delta \geq \Delta_{\min}.$

\vspace*{0.2cm}
Turning next to Theorems \ref{th:k_IRS_finite} and \ref{prop:k_FS_finite}, 
an $n$-length $k$-IRS code or a
fixed-set sampling block code can be viewed as restrictions of a $k$-MRS code.
Specifically, in Theorem \ref{th:k_IRS_finite},
for a $k$-IRS code of rate $R$
with $P_{S_{t}}, \ g(\theta) = \theta_2 $   
instead of $P_{S_{t}|X_{\cM t} }, \ g(\theta) = \theta$ (for a $k$-MRS),
the hypothesis of Lemma \ref{l:markov_single_letter} holds. 
Denote $\mathbbm{E}[d(X_{B}^{n}, Y_{B}^{n})|\theta_2 = \tau_2]$ 
by $\Delta_{\tau_2}, \ \tau_2 \in \Theta_2$.
Then, the pmfs in \eqref{eq:conv_joint_pmf} satisfy
\begin{align} \label{eq:conv_joint_pmf_IRS}
 P_{U \theta  X_{\cM} S Y_{B}} =  P_{U} \mu_{\theta} 
 P_{X_{\cM}|\theta} P_{S| U} P_{Y_{B}|S X_{S}  \theta U}.
\end{align}
The counterpart of \eqref{eq:conv_finite_eq5} is
\begin{align}
 R & \geq \underset{   P_{Y_{B}|S X_{S} U, \theta_2 = \tau_2} \atop 
 \mathbbm{E}[d(X_{B}, Y_{B})| \theta_2 = \tau_2] \leq \Delta_{\tau_2} } \min
 I(X_{S} \wedge Y_{B} |S, U, \theta_2 = \tau_2) \\
 & = \underset{   P_{Y_{B}|S X_{S} U, \theta_2 = \tau_2} \atop 
 \mathbbm{E}[d(X_{B}, Y_{B})| \theta_2 = \tau_2] \leq \Delta_{\tau_2} } \min
 \sum \limits_{A  , u } P_{S}(A) P_{U|S}(u|A) I(X_{A} \wedge Y_{B} 
 |S = A, U = u, \theta_2 = \tau_2),
\end{align}
noting from \eqref{eq:conv_joint_pmf_IRS} that $P_{U|S, \theta_2} = P_{U|S}.$ 
Using the convexity of the mutual information terms above with respect
to $P_{Y_{B}|S X_{S} U \theta_2}$, we get 
\begin{align}
 R  & \geq  \underset{   P_{Y_{B}|S X_{S} U, \theta_2 = \tau_2} \atop 
 \mathbbm{E}[d(X_{B}, Y_{B})| \theta_2 = \tau_2] \leq \Delta_{\tau_2} } \min  \sum \limits_{A  } P_{S}(A) I(X_{A} \wedge Y_{B} |S = A,  \theta_2 = \tau_2) \\
 & =   \rho_{\imath}^{\cB}(\Delta_{\tau_2}, P_{S}, \tau_2). \label{conv:finite_eq4}
\end{align}
Since \eqref{conv:finite_eq4} holds for every $\tau_2 \in \Theta_2$ 
\begin{align}
 R  & \geq   \underset{\tau_2 \in \Theta_2} \max \ \rho_{\imath}^{\cB}(\Delta_{\tau_2}, P_{S}, \tau_2) \\ 
 & \geq \underset{  P_{S}, \{ \Delta_{\tau_2}, \ \tau_2 \in \Theta_2 \} \atop \mathbbm{E}[\Delta_{\theta_2}] \leq \Delta  } \min \ \underset{\tau_2 \in \Theta_2} \max \ \rho_{\imath}^{\cB}(\Delta_{\tau_2}, P_{S}, \tau_2) \\
 & = R_{\imath}(\Delta),
\end{align}
i.e., $R \geq R_{\imath}(\Delta), \ \Delta \geq \Delta_{\min}, $ completing the converse
proof of Theorem \ref{th:k_IRS_finite}.

\vspace*{0.2cm}
In a manner analogous to a $k$-IRS, in Theorem \ref{prop:k_FS_finite} 
for a fixed-set sampler the hypothesis of Lemma \ref{l:markov_single_letter} holds with $P_{S_{t}} =  \mathbbm{1}(S_t = A), \ g(\theta) = \theta_1$. Defining $ \Delta_{\tau_1} \triangleq \mathbbm{E}[d(X_{B}^{n}, Y_{B}^{n})|\theta_1 = \tau_1], \ \tau_1 \in \Theta_1$,  the counterpart of the right-side of
\eqref{conv:finite_eq2} reduces to $ \underset{\tau_1 \in \Theta_1} \max \ \rho_{A}^{\cB}(\Delta_{\tau_1}, \tau_1).$ It then follows that 
$$R \geq \underset{ \{ \Delta_{\tau_1}, \ \tau_1 \in \Theta_1  \}  \atop \mathbbm{E}[\Delta_{\theta_1}] \leq \Delta} \min \ \underset{\tau_1 \in \Theta_1} \max \ \rho_{A}^{\cB}(\Delta_{\tau_1},\tau_1), \qquad \Delta \geq \Delta_{\min} $$ providing the converse proof for Theorem \ref{prop:k_FS_finite}.

\vspace*{0.3cm}

In the nonBayesian setting, the analog of Lemma \ref{l:markov_single_letter}
is obtained similarly with $C=c, \ g(C) = g(c),$ and \eqref{eq:markov_lemma_eq1}--\eqref{eq:markov_lemma_eq4}
expressed in terms of appropriate conditional pmfs. The converse proofs for a 
$k$-MRS, $k$-IRS and $k$-FS are obtained as above but by excluding the 
outer minimizations over $\{ \Delta_{\tau}, \tau \in \Theta \}$, $\{ \Delta_{\tau_2}, \tau_2 \in \Theta_2 \}$
and $\{ \Delta_{\tau_1}, \tau_1 \in \Theta_1 \}$, respectively.

\qed


\section{Discussion}


Our formulation of universality requires optimum sampling rate distortion
performance when the ``true'' underlying pmf of the DMMS belongs to a finite family
$\cP = \{ P_{X_{\cM}|\theta = \tau},  \  \tau \in \Theta\}$. The assumed
finiteness of $\Theta$ affords two benefits in addition to mathematical ease:
(i) simple proofs of estimator consistency uniformly over $\Theta_1, \ \Theta_2$ 
or $\Theta$; and 
(ii) rate-free conveyance of corresponding estimates ${\widehat \tau}_1, \ {\widehat \tau}_2$
or ${\widehat \tau}$ to the decoder.
General extensions to the case when $\Theta$ is an infinite set (countable or uncountable)
remain open.

\vspace*{0.1cm}

Unlike for a $k$-IRS, the assumption in a $k$-MRS that the decoder is informed of the 
sampling sequence $S^n$ plays an important role. Specifically, embedded information
regarding $X_{\cM}^n$ is conveyed implicitly to the decoder through $S^n$. Also, as
a side-benefit, the decoder can replicate the estimate of $\theta$ formed by the encoder 
based on $S^n$ alone, obviating the need for  explicitly transmitting it.
However, if the decoder were denied a knowledge of $S^n$, what is the USRDf?
This question, too, remains unanswered.

\vspace*{0.2cm}

Underlying our  achievability proofs of Theorems \ref{th:k_IRS_finite} and \ref{th:k_MRS_finite}
for a $k$-IRS and $k$-MRS, are schemes for distribution-estimation based on $(S^n,X_S^n).$
A distinguishing feature from classical estimation settings is the 
additional degree of (spatial) freedom in the choice of
the sampling sequence $S^n$. This motivates questions of the following genre: How
should $S^n$, consisting of (possibly different) $k$-sized subsets, be chosen to
form ``best'' estimates of the underlying joint pmf? How does the degree of the allowed
dependence of $S^n$ on $X_{\cM}^n$ affect estimator performance? For instance, our 
choice of sampling sequence and estimation procedure in the achievability proof of 
Theorem \ref{th:k_MRS_finite} is a simple starting point. How must we devise 
{\it efficient} sampling mechanisms to exploit an implicit embedding of DMMS realization
in the sampler output? These questions are of
independent interest in statistical learning theory.

\vspace*{0.1cm}

\renewcommand\baselinestretch{0.9}
{\small
\providecommand{\bysame}{\leavevmode\hbox to3em{\hrulefill}\thinspace}

}

\vspace*{0.3cm}

\appendix

\section{Appendices}



\noeqref{eq:MRS_rate_ac} \noeqref{eq:MRS_dist_ac}

\subsection{Proof of (\ref{eq:MRS_rate_ac}) and (\ref{eq:MRS_dist_ac})} \label{app:MRS_rate_dis}
For the code formed by concatenating $(f, \varphi)$ for each $u \in \cU$, the rate is
\begin{align}
&  \overset{\sim } \leq  \underset{\tau \in \Theta} \max  
\sum \limits_{u \in \cU} P_{U}(u) \frac{1}{n'}  \sum \limits_{i=1}^{M_{k}} \log J_{A_{i}}^{u, \tau}   \\
 & \leq  \underset{\tau \in \Theta} \max \sum \limits_{u \in \cU} P_{U}(u) \Big ( \sum \limits_{i=1}^{M_{k}} \frac{|\nu_{A_{i}}^{u, \tau}|}{n} \frac{1}{|\nu_{A_{i}}^{u, \tau}|}  \log J_{A_{i}}^{u, \tau} \Big) \\
 &   \leq  \underset{\tau \in \Theta} \max \sum \limits_{u \in \cU} P_{U}(u) \Big (  \sum \limits_{i=1}^{M_{k}} P_{S|U \theta}(A_i|u, \tau)  \Big (  I(X_{A_{i}} \wedge Y_{B} | S = A_{i}, U=u, \theta = \tau ) \! + \! \frac{\ep'}{2} \Big )  \Big ) , \quad \text{by } \eqref{eq:MRS_Bayes_fin_ac_rateA}  \\
 & \leq \underset{\tau \in \Theta} \max   \sum \limits_{u \in \cU} P_{U}(u)  \ I(X_{S} \wedge Y_{B}|S, U=u , \theta = \tau) + \ep'\\
 &\leq R_{m}(\Delta) + \ep, \label{eq:1MRS_Bayes_fin_ac_eq0}
\end{align}
for all $n$ large enough. 

\vspace*{0.2cm}

For each $U=u,$ let $\Delta_{u} \triangleq \sum \limits_{\tau \in \Theta, \ A_{i} \in \cAk} \mu_{\theta}(\tau) P_{S|U\theta}(A_{i}|u,\tau) \Delta_{A_{i}, u, \tau} $. Denoting the output of the decoder by $Y_{B}^{n' },$
we get
\begin{align}
 \mathbbm{E}[d(X_{B}^{n'}, Y_{B}^{n'})] & \leq P( {\widehat \tau}_{N} \neq \theta  ) d_{\max}  +  \mathbbm{E}[ \mathbbm{1}({\widehat \tau}_{N} = \theta)d(X_{B}^{n'}, Y_{B}^{n'})] \\
 & \leq P( {\widehat \tau}_{N} \neq \theta  ) d_{\max}  +  
 P( S^{\gamma} \notin {\cal T}^{(n)}(\ep',{\widehat \tau}_{N})  ) d_{\max} \\
 & \quad +  \mathbbm{E} \big [\mathbbm{E}[\mathbbm{1}({\widehat \tau}_{N} = \theta, S^{\gamma} \in {\cal T}^{(n)}(\ep',{\widehat \tau}_{N}))d(X_{B}^{n'}, Y_{B}^{n'})| S^{\gamma}, \theta] \big ] \label{eq:1MRS_Bayes_fin_ac_eq1} \\
 & \leq \mathbbm{E}[\Delta_{S, U , \theta} | U = u] + \ep \label{eq:1MRS_Bayes_fin_ac_eq2} \\ 
 & = \Delta_{u} + \ep 
\end{align}
for all $n, N$ large enough, where the previous inequality is shown below. Then, expected
distortion for the code formed by concatenating $(f, \varphi)$ for each $u \in \cU$, is
\begin{align}
 \substack{\sim \\ \leq \\\ } \ \mathbbm{E}[\Delta_{U}] + \ep \leq \Delta + \ep.
\end{align}
It remains to show \eqref{eq:1MRS_Bayes_fin_ac_eq2}. Now, \eqref{eq:1MRS_Bayes_fin_ac_eq2} follows
from the following: In \eqref{eq:1MRS_Bayes_fin_ac_eq1}, for each
$\tau \in \Theta$ and  $s^{n} \in {\cal T}^{(n)}({\ep'},{\widehat \tau}_{N})$,
\begin{align}
 \mathbbm{E}[ \mathbbm{1}( {\widehat \tau}_{N} = \theta  ) d(X_{B}^{n' }, & Y_{B}^{n'})  | S^{\gamma} = s^{n}, \theta = \tau ] \\
 & =  \mathbbm{E} \Big [  \frac{ \mathbbm{1}( {\widehat \tau}_{N} = \theta )}{n'} \sum \limits_{t \in \mu} d(X_{B t}, Y_{B t}) + \frac{ \mathbbm{1}( {\widehat \tau}_{N} = \theta )}{n'} \sum \limits_{t \in \gamma} d(X_{B t}, Y_{B t}) \big | S^{\gamma} = s^{n}, \theta = \tau \Big ] \\
 & \leq  \frac{ N'}{n'}d_{\max}  +  \frac{1}{n}   
 \mathbbm{E}\Big [\sum \limits_{i=1}^{M_{k}} \sum \limits_{t \in \gamma_{s^{n}}(A_{i}) \setminus \nu_{A_{i}} }d(X_{B {t}}, Y_{B {t}})|S^{\gamma} = s^{n},   \theta = \tau \Big]  \\
 & \ \ +  \sum \limits_{i=1}^{M_{k}}\mathbbm{E} \Big [  \frac{|\nu_{A_{i}}|}{n}   \mathbbm{1}( {\widehat \tau}_{N} = \theta ) d (X_{B}^{\nu_{A_{i}}}, \varphi_{A_{i}}^{ \theta }( f_{A_{i}}^{ \theta } (X_{A_{i}}^{\nu_{A_{i}}}) ) ) \big | S^{\nu_{A_{i}}} = A_{i}^{\nu_{A_{i}}},  \theta = \tau \Big ] \\
 & \leq \frac{ N'}{n'}d_{\max}  +  M_{k} \ep' d_{\max} + \sum \limits_{i=1}^{M_{k}} P_{S|U \theta}(A_{i}|u,\tau) \Big ( \Delta_{A_{i},u, \tau} + \frac{\ep'}{4} \Big ), \qquad \text{by }  \eqref{eq:MRS_Bayes_fin_ac_distA} \\
 & \leq \mathbbm{E}[\Delta_{S, U , \theta} | U = u, \theta = \tau]  +   M_{k} \ep' d_{\max} + \frac{ N'}{n'}d_{\max} + \frac{\ep'}{4} \\
 &\leq \mathbbm{E}[\Delta_{S, U , \theta} | U = u, \theta = \tau] + \ep,  
\end{align}
for all $n$ large enough and $\ep'$ chosen appropriately.
\qed


\vspace*{0.2cm}

\subsection{Proof of Proposition \ref{prop:MRS_fin_opt_samp} } \label{app:prop1}

\noindent First, for the Bayesian setting,  by Theorem \ref{th:k_MRS_finite}, the claim entails showing that
\begin{align}
 \underset{  P_{U}, P_{S|X_{\cM} U}, \{ \Delta_{\tau},  \ \tau \in \Theta \} \atop 
  \mathbbm{E}[\Delta_{\theta}] \leq \Delta}
  \min \ & \underset{\tau \in \Theta} \max \  \underset{P_{Y_{B}|S X_{S} U, \theta = \tau} \atop \mathbbm{E}[d(X_{B}, Y_{B})|\theta = \tau] \leq \Delta_{\tau}} \min \ I(X_{S} \wedge Y_{B} | S, U , \theta = \tau)  \label{eq:MRS_condnl_pt_mass_eq1}  \\
&  = \underset{  P_{U}, \delta_{w }, \{ \Delta_{\tau},  \ \tau \in \Theta \} \atop 
  \mathbbm{E}[\Delta_{\theta}] \leq \Delta}
  \min \ \underset{\tau \in \Theta} \max  \  \underset{P_{Y_{B}|S X_{S} U, \theta = \tau} \atop \mathbbm{E}[d(X_{B}, Y_{B})|\theta = \tau] \leq \Delta_{\tau}} \min \ I(X_{S} \wedge Y_{B} | S, U , \theta = \tau),  \label{eq:MRS_condnl_pt_mass_eq2}
 \end{align}
 for $\Delta_{\min} \leq \Delta \leq \Delta_{\max}$.
 Denote the expressions in \eqref{eq:MRS_condnl_pt_mass_eq1} and 
 \eqref{eq:MRS_condnl_pt_mass_eq2} by $q(\Delta)$ and $r(\Delta),$ respectively.
 Now, from the conditional version of Tops\o e's identity \cite[Lemma 8.5]{CsiKor11},
 observe that $q(\Delta)$ equals
\begin{align}
& \underset{  P_{U}, P_{S|X_{\cM} U}, \{ \Delta_{\tau},  \ \tau \in \Theta \} \atop 
  \mathbbm{E}[\Delta_{\theta}] \leq \Delta}
  \min \ \underset{\tau \in \Theta} \max \ 
  \underset{ P_{Y_{B}|S X_{S} U, \theta = \tau} \atop 
  \mathbbm{E}[d(X_{B}, Y_{B})|\theta = \tau] \leq \Delta_{\tau}}
  \min \ \underset{ Q_{Y_{B}|S U , \theta = \tau}} \min \  
  D \left ( P_{Y_{B}|SX_{S} U, \theta = \tau } \big | \big | Q_{Y_{B}| S U, \theta = \tau}
\big | P_{ S X_{S} U | \theta = \tau} \right ) \label{eq:MRS_condnl_pt_mass_eq6}.
\end{align}
Note that the inner max and min can be interchanged in \eqref{eq:MRS_condnl_pt_mass_eq6}.
Denoting 
$
  D \left ( P_{Y_{B}|SX_{S} U , \theta = \tau } \big | \big | Q_{Y_{B}| S U, \theta = \tau}
\big | P_{ S X_{S} U | \theta = \tau} \right )
$
by $D_{\tau}, \ \tau \in \Theta $, we write \eqref{eq:MRS_condnl_pt_mass_eq6} as
\begin{align}
& \underset{  P_{U}, P_{S|X_{\cM} U}, \{ \Delta_{\tau},  \ \tau \in \Theta \} \atop 
  \mathbbm{E}[\Delta_{\theta}] \leq \Delta}
  \min \  \underset{ P_{Y_{B}|S X_{S} U \theta }, Q_{Y_{B}|S U , \theta = \tau}
  \atop \mathbbm{E}[d(X_{B}, Y_{B})|\theta = \tau] \leq \Delta_{\tau}, \ \tau \in \Theta}
  \min \ \underset{\tau \in \Theta} \max  \ D_{\tau} 
  \\
& \quad  = \underset{  P_{U}, P_{S|X_{\cM} U} , P_{Y_{B}|S X_{S} U \theta},
Q_{Y_{B}|S U , \theta = \tau} \atop \mathbbm{E}[d(X_{B}, Y_{B})] \leq \Delta}
\min \ \underset{\tau \in \Theta} \max \  D_{\tau} 
\\
& \quad = \underset{
  \begin{subarray}{c}
  t, P_{U}, P_{S|X_{\cM} U} , P_{Y_{B}|S X_{S} U \theta}, Q_{Y_{B}| S U, \theta = \tau} \\
  D_{\tau} \leq t, \ \tau \in \Theta \\
  \mathbbm{E}[d(X_{B}, Y_{B}) ] \leq \Delta     
  \end{subarray}
} \min \ t,   \label{eq:MRS_condnl_pt_mass_eq3}
\end{align}
which is the epigraph form. Also,
$r(\Delta)$ can be expressed in a similar manner. 
Based on \eqref{eq:MRS_condnl_pt_mass_eq3}, we define 
$ G_{q}(\alpha, \{ \lambda_{\tau}, \tau \in \Theta \} )$ and 
$ G_{r}(\alpha, \{ \lambda_{\tau}, \tau \in \Theta \} )$ in terms 
of the Lagrangians of $q(\Delta)$ and $r(\Delta)$, respectively, in a standard way.
 
Specifically, $G_{q}( \alpha, \{ \lambda_{\tau}, \ \tau \in \Theta \} )$ 
\begin{align}
& = \underset{t, P_{U}, P_{S|X_{\cM} U} \atop P_{Y_{B}|S X_{S} U \theta},
Q_{Y_{B}|S  U \theta}} \min
\hspace*{0.4cm}     t +  \sum \limits_{\tau \in \Theta} \lambda_{\tau} 
(D_{\tau} - t)
      + \alpha \mathbbm{E}\left[ d(X_{B}, Y_{B}) \right]  \\ 
  & = \underset{t, P_{U}, P_{S|X_{\cM} U} \atop P_{Y_{B}|S X_{S} U \theta},
  Q_{Y_{B}|S  U \theta}} \min
\hspace*{0.4cm}     t(1 - \sum \limits_{\tau \in \Theta} \lambda_{\tau}) +  
\sum \limits_{\tau \in \Theta} \lambda_{\tau} D_{\tau}
      + \alpha \mathbbm{E}\left[ d(X_{B}, Y_{B}) \right]  \\ \vspace*{2cm}
 & = \begin{cases}
     \underset{ P_{U}, P_{S|X_{\cM} U} \atop P_{Y_{B}|S X_{S} U \theta},
     Q_{Y_{B}|S  U \theta}} \min \ \sum \limits_{\tau \in \Theta} \lambda_{\tau} D_{\tau}  
   + \alpha \mathbbm{E}\left[ d(X_{B}, Y_{B}) \right], \ \text{ if }
      \sum \limits_{\tau \in \Theta} \lambda_{\tau} = 1 \\
      - \infty, \ \hspace*{7 cm} \text{otherwise}. \label{eq:app_eq1}
     \end{cases}
\end{align}

\noindent Let $P_{\tau} \triangleq P_{X_{\cM}|\theta = \tau}   $. 
When $ \sum \limits_{\tau \in \Theta} \lambda_{\tau} = 1,$ from \eqref{eq:app_eq1},
$G_{q}(\alpha, \{ \lambda_{\tau}, \ \tau \in \Theta \}  )$ equals
\begin{align}
 &  \underset{ P_{U}, Q_{Y_{B}| S U \theta}, \atop  P_{ Y_{B} | S X_{S}U \theta} }
 \min \!  \sum \limits_{u,x_{\cM}} P_{U  }(u )  \underset{ P_{S|X_{\cM}U}   } 
 \min \sum \limits_{s \in \cAk} P_{S|  X_{\cM} U}(s|x_{\cM},u) \times \\
 & \! \Bigg ( \! 
  \mathbbm{E} \Big [ \! \sum \limits_{\tau \in \Theta} \lambda_{\tau}
  P_{\tau}(x_{\cM} ) \!    \log \frac{ P_{Y_{B}|S X_{S} U \theta} 
  (Y_{B} | s,x_{s},u, \tau ) }{Q_{Y_{B}|SU \theta} (Y_{B}| s, u,\tau) }  \!  +  \!
 \alpha \sum \limits_{\tau \in \Theta} \mu_{\theta}(\tau) P_{\tau}(x_{\cM} )  
 d(x_{B}, Y_{B} ) \Big | S = s, X_{S} = x_{s}, U = u, \theta = \tau \Big ] \! \Bigg ),   
\end{align}
where the expectation above is with respect to $P_{ Y_{B} | S = s,
X_{S} = x_{s},U=u, \theta = \tau} $. Noting that the term $\Big(\cdots  \Big )$
above is a function of $s, x_{\cM}, u,$ we get
\begin{align}
G_{q}( \alpha, & \{ \lambda_{\tau}, \ \tau \in \Theta \}  )   \\
& = \underset{ P_{U}, Q_{Y_{B}| S U \theta}  \atop  P_{ Y_{B} | S X_{S}U \theta} } 
\min \   \sum \limits_{u,x_{\cM}} P_{U  }(u )  \underset{ s \in \cAk   } \min  
 \Bigg (  \mathbbm{E} \Big [ \sum \limits_{\tau \in \Theta} \lambda_{\tau} 
 P_{\tau}(x_{\cM} ) \log \frac{ P_{Y_{B}|S X_{S} U \theta} (Y_{B} | s,x_{s},u, \tau ) }{Q_{Y_{B}|SU \theta} (Y_{B}| s, u,\tau) }  \\ & \hspace*{5.7 cm} +  
 \alpha \sum \limits_{\tau \in \Theta} \mu_{\theta}(\tau) P_{\tau}(x_{\cM} ) 
 d(x_{B}, Y_{B} ) \Big | S = s, X_{S} = x_{s}, U = u, \theta = \tau \Big ] \Bigg ) \\
 & = \underset{ P_{U}, Q_{Y_{B}| S U \theta}  \atop  P_{ Y_{B} | S X_{S}U \theta} }
 \min \   \sum \limits_{u,x_{\cM}} P_{U  }(u )  \ \underset{ \delta_{w (\cdot, \cdot)}   } \min \sum \limits_{s \in \cAk} \delta_{w(x_{\cM},u)}(s)
 \Bigg (  \mathbbm{E} \Big [ \sum \limits_{\tau \in \Theta} \lambda_{\tau}
 P_{\tau}(x_{\cM} ) \log \frac{ P_{Y_{B}|S X_{S} U \theta} (Y_{B} | s,x_{s},u, \tau ) }
 {Q_{Y_{B}|SU \theta} (Y_{B}| s, u,\tau) }  \\ & \hspace*{5.7 cm} +  
 \alpha \sum \limits_{\tau \in \Theta} \mu_{\theta}(\tau)  P_{\tau}(x_{\cM} )  
 d(x_{B}, Y_{B} ) \Big | S = s, X_{S} = x_{s}, U = u, \theta = \tau \Big ] \Bigg ) \\
& = \underset{ P_{U}, Q_{Y_{B}|S  U \theta}\atop 
P_{Y_{B}|S X_{S} U \theta},\delta_{w } } \min \sum \limits_{\tau \in \Theta} 
\lambda_{\tau} D \left ( P_{Y_{B}|SX_{S} U, \theta = \tau } \big | \big 
| Q_{Y_{B}| S U, \theta = \tau}  \big | P_{ S X_{S} U | \theta = \tau} \right )  
+ \alpha \mathbbm{E}\left[ d(X_{B}, Y_{B}) \right] \\
& = G_{r}(\alpha, \{ \lambda_{\tau}, \ \tau \in \Theta \} ).
\end{align}

Since $q(\Delta)$ and $r(\Delta)$
are convex in $\Delta,$ they can be expressed in terms of their respective Lagrangians as 
\begin{align}
 q(\Delta) = \underset{\alpha \geq 0, \ \{ \lambda_{\tau} \geq 0, 
 \ \tau \in \Theta \} } \max G_{q}( \alpha, \{ \lambda_{\tau}, 
 \tau \in \Theta \}) - \alpha \Delta  \ \text{    and     } \ 
 r(\Delta) = \underset{\alpha \geq 0, \ \{ \lambda_{\tau} \geq 0, \ 
 \tau \in \Theta \}  } \max G_{r}(\alpha, \{ \lambda_{\tau}, \tau \in \Theta \} )
 - \alpha \Delta.  \label{eq:MRS_condnl_pt_mass_eq4}
\end{align}

Thus, 
\begin{align}
q(\Delta)  =   \underset{ \alpha \geq 0, \ \{ \lambda_{\tau} \geq 0, 
\ \tau \in \Theta \}   } \max  G_{q}(\alpha, \{ \lambda_{\tau}, \ 
\tau \in \Theta \} ) - \alpha \Delta & = \underset{ \alpha \geq 0, \
\{ \lambda_{\tau} \geq 0, \ \tau \in \Theta \} \atop \sum \limits_{\tau \in \Theta} 
\lambda_{\tau} = 1  } \max  G_{q}(\alpha, \{ \lambda_{\tau}, \ \tau \in \Theta \} ) - \alpha \Delta \\
& = \underset{ \alpha \geq 0, \  \{ \lambda_{\tau} \geq 0, \ \tau \in \Theta \} 
\atop \sum \limits_{\tau \in \Theta} \lambda_{\tau} = 1  } \max 
G_{r}(\alpha, \{ \lambda_{\tau}, \ \tau \in \Theta \}) - \alpha \Delta \\
& = r(\Delta),
\end{align}
upon observing that the maxima in \eqref{eq:MRS_condnl_pt_mass_eq4} are 
attained when $\sum \limits_{\tau \in \Theta} \lambda_{\tau} = 1$.

\qed


\subsection{ Proof of Lemma \ref{l:convexity}} \label{app:lemma_std_props}

Clearly, for each $\tau_1 \in \Theta_1$,
$\rho_{A}^{\cB}(\delta, \tau_1)$ and $\rho_A^{n \cB}(\delta, \tau_1)$ are finite-valued 
and, hence, so are the right-sides of \eqref{eq:Bayesian_kfs} and \eqref{eq:nonBayesian_kfs}. 
Also, they are also nonincreasing in $\Delta$.
The convexity of the right-sides of \eqref{eq:Bayesian_kfs}
and \eqref{eq:nonBayesian_kfs} follows from the convexity of $\rho_{A}^{\cB}(\delta, \tau_1)$
and $\rho_{A}^{n\cB}(\delta, \tau_1)$ in $\delta$ along with a standard argument shown below;
continuity for $\Delta > \Delta_{\min}$ is a consequence. Continuity at $\Delta_{\min}$ 
holds, for instance, as in (\cite{CsiKor11}, Lemma 7.2). 
\noindent The claimed properties of the right-sides of \eqref{eq:USRDf_IRS_Bayesian}, \eqref{eq:USRDf_IRS_nonBaye_fin},
\eqref{eq:USRDf-MRS-Bayesian-finite} and \eqref{eq:USRDf-MRS-nonBayesian-finite} follow 
in a similar manner.

\vspace*{0.2cm}

The convexity of the right-side of \eqref{eq:Bayesian_kfs} can be shown explicitly as 
follows. Let $\tau_1(1)$ and $\tau_1(2)$ attain the maximum in \eqref{eq:Bayesian_kfs}
at $\Delta = \Delta_1$ and $\Delta = \Delta_2$, respectively, where $\Delta_1 < \Delta_2$.
The corresponding minimizing $\{ \Delta_{\tau_1}, \tau_1 \in \Theta_1 \}$ are denoted by 
$\{ \Delta_{\tau_1}^1, \ \tau_1 \in \Theta_1 \}$ and $\{ \Delta_{\tau_1}^2, \ \tau_1 \in \Theta_1 \}$,
respectively.
For any $ 0 < \alpha < 1$,  
for   $i = 1, \ldots, |\Theta_1|$
\begin{align}
 \alpha R_{A}(\Delta_1) + (1-\alpha) R_{A}(\Delta_2)
 & = \alpha \rho_{A}^{\cB}(\Delta_{\tau_1(1)}^1,\tau_1(1)) + (1-\alpha) \rho_{A}^{\cB}(\Delta_{\tau_1(2)}^2,\tau_1(2)) \\
 & \geq \alpha \rho_{A}^{\cB}(\Delta_{\tau_1(i)}^1,\tau_1(i)) + (1-\alpha) \rho_{A}^{\cB}(\Delta_{\tau_1(i)}^2,\tau_1(i))  \\
 & \geq \rho_{A}^{\cB} ( \alpha  \Delta_{\tau_1(i)}^1 + (1-\alpha) \Delta_{\tau_1(i)}^2, \tau_1(i) ) , \label{eq:l_convex_eq1} 
\end{align}
where the inequality above
follows by Remark (iii) preceding Theorem \ref{prop:k_FS_finite} in Section \ref{s:Results}.
Now, \eqref{eq:l_convex_eq1}
holds for every $i =  1, \ldots, |\Theta_1|,$ hence 
\begin{align}
 \alpha R_{A}(\Delta_1) + (1-\alpha) R_{A}(\Delta_2) 
 &  \geq \underset{i} \max \  \rho_{A}^{\cB} ( \alpha  \Delta_{\tau_1(i)}^1 + (1-\alpha) \Delta_{\tau_1(i)}^2, \tau_1(i) ) \\
 & \geq  \underset{ \{\Delta_{\tau_1}, \tau_1 \in \Theta_1 \} \atop \mathbbm{E}[ \Delta_{\theta_1}] \leq \alpha \Delta_1 + (1-\alpha) \Delta_2 } \min \underset{\tau_1 \in \Theta_1} \max \ \rho_{A}^{\cB}(\Delta_{\tau_1}, \tau_1) \\
 & =  R_{A}( \alpha \Delta_1 + (1-\alpha) \Delta_2). 
\end{align}

\vspace*{-0.2cm}

\qed


\begin{thebibliography}{20}

\bibitem{Ber71}
T. Berger,
{\it Rate Distortion Theory: A Mathematical Basis for Data Compression},
Prentice Hall, Englewood Cliffs, NJ, 1971.


\bibitem{Ber78}
T.~Berger, 
ÒMultiterminal source coding,Ó in {\it The Information Theory
Approach to Communications}, G. Longo, Ed. Vienna/New York:
Springer-Verlag, 1978, vol. 229, CISM Courses and Lectures, pp.
171--231.

 


\bibitem{BodNar17}
V.~P.~Boda and P.~Narayan, ``Sampling rate distortion,'' 
{\it IEEE Trans. Inform. Theory}, vol.~63, no.~1, pp. 563--574, Jan. 2017.


\bibitem{CoverThomas}
T.~M.~Cover and J.~A.~Thomas, {\it Elements of Information Theory}, John Wiley \& Sons, 2012.



\bibitem{CsiKor11}
I.~Csisz\'ar and J.~K\"{o}rner, {\it Information Theory: Coding Theorems for Discrete 
Memoryless Systems},  Cambridge University Press, 2011.


\bibitem{DemboWeiss03}
A.~Dembo and T.~Weissman, 
``The minimax distortion redundancy in noisy source coding,''
{\it IEEE Trans. Inform. Theory}, vol.~49, no.~11, 
pp.~3020--3030, Nov. 2003.


\bibitem{DobTsy62}
R.~L.~Dobrushin and B.~S.~Tsybakov, ``Information transmission with
additional noise," {\it IRE Trans. Inform. Theory}, vol.~IT-8, pp. 293--304, Sept. 1962.


\bibitem{GiriKum05}
A.~Giridhar and P.~Kumar, ``Computing and communicating functions
over sensor networks,'' {\it IEEE Journ. on Select. Areas in Commun.}, 
vol.~23, no.~4, pp. 755--764, April 2005.


\bibitem{IshKunRam03}
P.~Ishwar, A.~Kumar and K.~Ramachandran,
``On distributed sampling in dense sensor networks: a ``bit conservation principle,"
{\it International Symposium on Information Processing in
Sensor Networks} (IPSN), Palo Alto, CA, April 2003.


\bibitem{KashLasXiaLiu05}
A.~Kashyap, L.~A. Lastras-Montano, C.~Xia and L.~Zhen, ``Distributed source coding in 
dense sensor networks," {\it Proceedings of Data Compression Conference, 2005}, 
pp. 13--22,  March 29-31, 2005.


\bibitem{KipnisGold16}
A.~Kipnis, A.~J.~Goldsmith, Y.~C.~Eldar and T.~Weissman,
``Distortion rate function of sub-Nyquist sampled Gaussian sources,"
{\it IEEE Trans. Inform. Theory}, vol.~62, no.~1, pp. 401--429, Jan. 2016.

\bibitem{KonTelVet12}
R.~L. Konsbruck, E.~Telatar and M.~Vetterli,
 ``On sampling and coding for distributed acoustic sensing," 
 {\it IEEE Trans. Inform. Theory}, vol.~58, no.~5, pp. 3198--3214, May 2012.


\bibitem{Linder00}
T.~Linder, ``On the training distortion of vector quantizers," 
{\it IEEE Trans. Inform. Theory}, vol.~46, no.~4, pp.~1617--1623, July 2000.

\bibitem{Linder02}
T.~Linder, ``Learning-theoretic methods in vector quantization," 
{\it Principles of nonparametric learning}, pp.~163--210, Springer Vienna, 2002.


\bibitem{LinLugZeg95}
T.~Linder, G.~Lugosi and K.~Zeger, ``Fixed-rate universal lossy 
source coding and rates of convergence for memoryless sources," 
{\it IEEE Trans. Inform. Theory}, vol.~41, no.~3, pp.~665--676, May 1995.


\bibitem{LinLugZeg97}
T.~Linder, G.~Lugosi and K.~Zeger, ``Empirical quantizer 
design in the presence of source noise or channel noise,'' 
{\it IEEE Trans. Inform. Theory}, vol.~43, no.~2, pp.~612--623,
March 1997.



\bibitem{LiuSimErk12}
 X.~Liu, O.~Simeone and E.~Erkip, ``Lossy computing of correlated sources with fractional 
 sampling," {\it Proceedings of the IEEE Information Theory Workshop (ITW), 2012},  
 pp. 232--236, Sept. 3-7, 2012.
 
 
\bibitem{NeuGrDav75}
D.~Neuhoff, R.~Gray and L.~Davisson, ``Fixed rate universal 
block source coding with a fidelity criterion," 
{\it IEEE Trans. Inform. Theory}, vol.~21, no.~5, pp.~511--523, Sept. 1975.

\bibitem{NeuPra11}
D.~L. Neuhoff and S.~S. Pradhan,
 ``Information rates of densely sampled Gaussian data," 
 {\it Proceedings of the IEEE International Symposium on 
 Information Theory Proceedings (ISIT), 2011}, pp. 2776--2780, July 31-Aug. 5, 2011.




\bibitem{NeuShi78}
D.~Neuhoff and P.~Shields, ``Fixed-rate universal codes for
Markov sources," {\it IEEE Trans. Inform. Theory}, vol.~24, 
no.~3, pp.~360--367, May 1978. 


 
 

\bibitem{ReeGas12}
G.~Reeves and M.~Gastpar,
 ``The sampling rate-distortion tradeoff for sparsity pattern recovery in compressed sensing," 
 {\it IEEE Trans. Inform. Theory}, vol.~58, no.~5, pp. 3065--3092, May 2012. 
 

 
\bibitem{Rissanen84}
J.~Rissanen, ``Universal coding, information, prediction and estimation,'' 
{\it IEEE Trans. Inform. Theory}, vol.~30, no.~4, pp.~629--636, July 1984.


 
\bibitem{WeiVet12}
C.~Weidmann and M.~Vetterli,
 ``Rate distortion behavior of sparse sources," 
 {\it  IEEE Trans. Inform. Theory}, vol.~58, no.~8, pp. 4969--4992, Aug. 2012.

\bibitem{Weiss01}
T.~Weissman, ``Universally attainable error exponents for 
rate-distortion coding of noisy sources,'' 
{\it IEEE Trans. Inform. Theory}, vol.~50, no.~6, pp.~1229--1246, June 2001.


\bibitem{WeissMer01}
T.~Weissman and N.~Merhav, ``Universal prediction of individual 
binary sequences in the presence of noise,'' 
{\it IEEE Trans. Inform. Theory}, vol.~47, no.~6, pp.~2151--2173, Sept. 2001.


\bibitem{WeissMer02a}
T.~Weissman and N.~Merhav, ``On limited-delay lossy coding and 
filtering of individual sequences,'' 
{\it IEEE Trans. Inform. Theory}, vol.~48, no.~3, pp.~721--733, March 2002.


 
\bibitem{WuVer10}
Y.~Wu and S.~Verd{\'u},
 ``R{\'e}nyi information dimension: fundamental limits of almost lossless analog compression," 
 {\it IEEE Trans. Inform. Theory}, vol.~56, no.~8, pp. 3721--3748, Aug. 2010.
 

\bibitem{YamIto80}
H.~Yamamoto and K.~Itoh,
 ``Source coding theory for multiterminal communication systems with a remote source,"
 {\it IEICE Trans.}, vol.~E63-E, no.~10, pp.~700--706, Oct. 1980.

\bibitem{ZhaSri10}
Y.~Zhang, A.~Srivastava and M.~Zahran, ``On-chip sensor-driven efficient thermal 
profile estimation algorithms,'' {\it ACM Trans. Des. Autom. Elec. Sys. (TODAES)}, vol.~15, no.~3,
p. 25, May 2010. 

\bibitem{Ziv72}
J.~Ziv, ``Coding of sources with unknown statistics--II: 
Distortion relative to a fidelity criterion,'' 
{\it IEEE Trans. Inform. Theory}, vol.~18, no.~3, pp.~389--394, May 1972. 

\bibitem{Ziv80}
J.~Ziv, ``Distortion-rate theory for individual sequences,'' 
{\it IEEE Trans. Inform. Theory}, vol.~26, no.~2, pp.~137--143, March 1980.
 
\end{thebibliography}
\end{document}